\documentclass[preprint,preprintnumbers,amsmath,amssymb]{revtex4-2}
\usepackage{graphicx}
\usepackage{graphics}
\usepackage{chngcntr}
\usepackage{bm}
\usepackage{epsfig}
\begin{document}

\title{Exclusive rare semileptonic decays of $B$ and $B_c$ mesons \\ in the relativistic independent quark model}

	\author{Kalpalata Dash$^{1}$\footnote{email address:kalpalatadash982@gmail.com}, P. C. Dash$^{1}$, R. N. Panda$^{1}$, Susmita Kar$^{2}$, N. Barik$^{3}$}
\address{$^{1}$ Department of Physics, Siksha $'O'$ Anusandhan Deemed to be University, Bhubaneswar-751030, India. \\$^{2}$ Department of Physics, Maharaja Sriram Chandra Bhanja Deo University,\\ Baripada-757003, India.\\ 
	$^{3}$ Department of Physics, Utkal University, Bhubaneswar-751004, India.}

\begin{abstract}
We investigate the exclusive rare semileptonic decays: $B(B_c)\to K(D_{(s)})l\bar{l}/\Sigma\nu_l\bar{\nu_l}$ ($l=\mu, \tau$) in the framework of relativistic independent quark (RIQ) model based on an average flavor independent confining potential in equally mixed scalar-vector harmonic form. The invariant weak form factors, parametrising the matrix elements between participating meson states are calculated in the parent meson rest frame. The momentum transfer dependence of the form factors is reliably determined in the whole accessible kinematical range: $q^2_{min}\leq 0\leq q^2_{max}$. Our predicted branching fractions for $B\to K\mu^+\mu^-/\tau^+\tau^-(\Sigma\nu_l\bar{\nu_l})$, $B_c\to D(D_s)\mu^+\mu^-$, $B_c\to D(D_s)\tau^+\tau^-$ and $B_c\to D(D_s)\Sigma\nu_l\bar{\nu_l}$, obtained in order of $10^{-7}(10^{-6})$,  $10^{-9}(10^{-7})$,  $10^{-9}(10^{-8})$ and  $10^{-8}(10^{-7})$, respectively are in reasonable agreement with other Standard Model predictions and Lattice QCD results. The averaged values of the lepton polarization asymmetries for decay modes are obtained as $\langle P_L(B\to K\mu^+\mu^-)\rangle$=-0.97, $\langle P_L(B_c\to D_{(s)}\mu^+\mu^-)\rangle$=-0.972, $\langle P_L(B\to K\tau^+\tau^-)\rangle$=-0.224, $\langle P_L(B_c\to D\tau^+\tau^-)\rangle$=-0.275 and $\langle P_L(B_c\to D_s\tau^+\tau^-)\rangle$=-0.194.  
\end{abstract}

\maketitle
\section{introduction}
The exclusive rare semileptonic $B_{(c)}$ decays are of special interest as they are governed by flavor-changing neutral current (FCNC), which are forbidden at the tree level in the Standard Model (SM) and proceed at the lowest order through one-loop diagram \cite{grinstein1989b, buras1995effective, misiak1993b, inami1981effects, buchalla1996weak, ali1991forward, kim1997b, aliev1998exclusive} involving at least one off-diagonal element of the Cabbibo-Kobayashi-Maskawa (CKM) matrix element $V_{tq}$ $(q=d, s, b)$. Being sensitive to the contribution of new intermediate particles that might appear in the loop, these decays are especially important as they potentially hint at New Physics (NP) beyond SM \cite{egede2010new, altmannshofer2009new}. They also provide a framework for precise determination of the CKM matrix element and leptonic decay constants of $D_{(s)}$ and $B_{(c)}$ mesons. 
\par The rarity of these decays, particularly involving lepton and neutrinos, presents significant experimental challenges due to their extremely low branching fractions (BFs). However, the first observed data on the BF: ${\cal B}(B\to K{l^+}{l^-})=(0.75^{+0.25}_{-0.21}\pm0.09)\times10^{-6}$, $l=e, \mu$ by Belle Collaboration \cite{abe2001observation} is made available in 2001. Since then a series of experimental probes made by Belle \cite{aubert2009direct, wei2009measurement, lees2012measurement}, BABAR \cite{aubert2006measurements, 
 lees2017search, choudhury2021test}, CDF \cite{aaltonen2009search, Pueschel:2010rm} and LHCb \cite{aaij2012measurement, aaij2013differential, aaij2014differential, aaij2014test, aaij2017measurement, lhcb2022test} Collaboration have yielded the observed data on number of rare decay properties such as the BFs; forward-backward asymmetry; direct CP, lepton flavor and isospin asymmetries; angular analysis and test of lepton-flavor universality etc. A comparison of existing observed data and theoretical predictions provides one of the most rigid constraints on the NP scenario \cite{hurth2010radiative}. The first-ever measurement of the BF by Belle II Collaboration: ${\cal B} (B^+\to K^+\nu_l\bar{\nu_l}) = (2.3\pm0.7)\times 10^{-5}$ \cite{adachi2024evidence}, which is nearly 3.5$\sigma$ higher than the SM prediction, raises intriguing possibilities for NP beyond SM. The data in $B_c$ sector is scant. However, with the advances in the Large Hadron Collider beauty experiments at LHC, there is a fair chance of around $5\times10^{10}$ $B_c$-events per year \cite{gouz2004prospects, PepeAltarelli:2008yyl}. This provides a good platform to explore the $B_c$ rare decays to pseudoscalar ($D, D_s$) and $l\bar{l}/\nu_l\bar{\nu_l}$.
\par The theoretical analysis of the rare weak decays is based on the effective electroweak Hamiltonian, derived from the leading and next-to-leading loop diagrams in the SM using the operator product expansion and renormalization group techniques \cite{buchalla1996weak, buras1995effective}. The QCD corrections to these decays due to hard gluon exchange, require to resum large logarithms which is done with the help of renormalization group methods. The operator product expansion allows to separate the perturbative short-distance effect encoded in the Wilson Coefficients of the effective Hamiltonian from the nonperturbative long-distance part encoded in the hadronic form factors. This part of the calculation is model-dependent since it involves the nonperturbative QCD. Therefore, it is important to have a reliable estimate of the hadronic form factors for the exclusive rare semileptonic $B_{(c)}$- decays and predict the decay properties within and beyond the SM.
\par The exclusive rare semileptonic $B_{(c)}$ meson decays have been studied by several theoretical approaches. Although we may not be able to list them all, some noteworthy among them include the relativistic constituent quark
model (RCQM) \cite{faessler2002exclusive, ebert2010rare}, light front quark model (LFQM) \cite{PhysRevD.65.094037, choi2002light, choi2010light}, light cone quark model (LCQM) \cite{wang2011b, dhiman2018study}, light cone sum rule (LCSR) \cite{ball2005new, khodjamirian2010charm, aliev1997light}, QCD sum rule (QCD SR) \cite{colangelo1996qcd, azizi2008analysis}, perturbative QCD (pQCD) \cite{wang2014semileptonic} and lattice QCD (LQCD) \cite{bobeth2007angular, bobeth2012decay, altmannshofer2012cornering, wang2012semileptonic, bobeth2013general, khodjamirian2013b, bouchard2013standard, bouchard2013rare, bailey2016b, parrott2023standard, hpqcd2023standard} approaches. In view of the potential prospects of recent experimental techniques and, in particular, the facilities available for the Large Hadron beauty experiments, dedicated to heavy flavor physics at LHC, the unmeasured exclusive rare semileptonic decays can hopefully be measured in the near future. The observed data in this sector \cite{abe2001observation, aubert2009direct, lees2012measurement, lees2017search, wei2009measurement, choudhury2021test, aaij2012measurement, aaij2013differential, aaij2014differential, aaij2014test, aaij2017measurement, lhcb2022test} can distinguish the predictions from several theoretical approaches on rare semileptonic $b$- flavored mesons. The prospects in the ongoing and future experiments in the heavy flavor sector and theoretical predictions available in the literature motivate us to analyse the exclusive rare semileptonic decays of $b$- flavored mesons in the framework of our relativistic independent quark (RIQ) model.
\par The RIQ model, developed by our group, has been applied in describing wide-ranging hadronic properties including static properties of hadrons \cite{barik1983octet, Barik:1986mq, barik1998elastic, barik1987and, barik1993weak} such as the elastic form factor, magnetic moment, charge radii of hadrons and the mass spectra etc. and several decay properties of hadrons including the semileptonic decays \cite{barik1996exclusive, barik1997exclusive, barik2009semileptonic, patnaik2020semileptonic, nayak2021lepton, nayak2022exclusive} of heavy-flavored mesons. In this paper, we would like to extend the applicability of RIQ model and study the decays $B(B_c)\to K(D_{(s)})l\bar{l}$ and $B(B_c)\to K(D_{(s)})\Sigma\nu_l\bar{\nu_l}$, where $l$ stands for the leptons $e$, $\mu$ or $\tau$. These decays are, in fact, governed by three independent invariant hadronic form factors ($f_+(q^2)$, $f_-(q^2)$ and $f_T(q^2)$). The form factors $f_+(q^2)$, $f_-(q^2)$ represent the hadronic matrix element of the vector-axial vector current and $f_T(q^2)$ represents the matrix element of the tensor current. The invariant weak decay form factors expressed as overlap integrals of the meson wave functions have been obtained in the corresponding calculation of mass spectra, and other static \cite{barik1983octet, Barik:1986mq, barik1998elastic, barik1987and, barik1993weak} and several decay properties including those Ref. \cite{barik1996exclusive, barik1997exclusive, barik2009semileptonic, patnaik2020semileptonic, nayak2021lepton, nayak2022exclusive} in the RIQ model framework in good agreement with the observed data and other SM predictions. In this model, the weak form factors are reliably extracted, where their dependence on the momentum transfer squared ($q^2$) is automatically encoded kinematical range. This is unlike in some other theoretical approaches available in the literature, where the form factors are determined first, with an endpoint normalization at minimum $q^2$ (maximum recoil) or maximum $q^2$ (minimum recoil) and then extrapolated to the entire kinematical region using some monopole/dipole/Gaussian ansatz, which makes the form factor estimation less reliable. In the present context, we would calculate the form factors ($f_+(q^2)$, $f_-(q^2)$ and $f_T(q^2)$) as well as BFs and longitudinal lepton polarization asymmetry (LPA) within the SM framework, using our RIQ model conventions. The LPA is an important asymmetry and a parity-violating observable \cite{hewett1996tau, kruger1996lepton, faessler2002exclusive, choi2010light}, which could be accessible in the LHCb experiment. This could be more accessible experimentally in $\tau$-channels; since it is proportional to the lepton mass in the SM. Various theoretical approaches \cite{fukae2000systematic, aliev2001new, aliev2001general, choi2002light, faessler2002exclusive, choi2010light}, have predicted this observable, using the general form of the effective Hamiltonian including all possible forms of interaction. In this paper, we would like to address these issues including the lepton mass effect in the rare semileptonic decay within our RIQ model framework. 
\par The paper is organized as follows. In Sec. II we discuss
the general formalism and kinematics for rare decays of $B(B_c)\to K(D_{(s)})l\bar{l}$ and $B(B_c)\to K(D_{(s)})\Sigma\nu_l\bar{\nu_l}$, which include a brief description of the effective Hamiltonian, invariant form factors and differential decay distribution, helicity amplitude and differential decay rate, and longitudinal lepton polarization asymmetry. In Section III, we extract the invariant transition form factors in the relativistic independent quark (RIQ) model. Our numerical analysis and discussion on the form factors, branching fractions and longitudinal polarization asymmetries are described in Section IV. Section V encompasses our Summary and Conclusion. A brief description of the RIQ model conventions, quark orbitals and momentum probability amplitude of constituent quark and antiquark in the RIQ model are given in the Appendix.
\section{General formalism and kinematics}
\subsection{Effective Hamiltonian}
The SM description of rare $b$-flavored meson decays is based on the low energy effective Hamiltonian, obtained by integrating out the degrees of freedom (the top quark and $W$ bosons). The effective Hamiltonians for $b\to q_il\bar{l}$ ($q_1=d, q_2=s$) and $b\to q_i\Sigma\nu_l\bar{\nu_l}$, renormalized at a scale $\mu=m_b$, given in \cite{buchalla2000phenomenology}, are 
\begin{eqnarray}
{\cal H}_{eff}^{l\bar{l}}=&&-\frac{4G_F}{\sqrt{2}} V_{tb}V_{t{q_i}}^*\sum_{i=1}^{10}C_i{\cal O}_i,\nonumber\\
{\cal H}^{\nu_l\bar{\nu}_l}_{eff}=&&-\frac{4G_F}{\sqrt{2}} V_{tb}V_{t{q_i}}^*\ C_L^{\nu_l}{\cal O}_L^{\nu_l},
\end{eqnarray}
where $G_F$ is the Fermi coupling constant, $V_{ij}$ is the Cabibbo-Kobayashi-Maskawa matrix elements, $C_i$ are the Wilson coefficients and ${\cal O}_i$ are the SM operator basis given in Ref \cite{buras1995effective, buchalla1996weak}. The operators that enter into the description of the $b\to q_il\bar{l}$ transitions are:
\begin{eqnarray}
    {\cal O}_7=&&\frac{e^2}{32\pi^2}m_b \ {(\bar{q_i}\sigma^{\mu\nu}(1+\gamma_{5}) b)}F_{\mu\nu},\nonumber\\
    {\cal O}_9=&&\frac{e^2}{32\pi^2} \ {(\bar{q_i}\gamma^\mu(1-\gamma_{5}) b)}(\bar{l}{\gamma_\mu }l),\nonumber\\
    {\cal O}_{10}=&&\frac{e^2}{32\pi^2} \ {(\bar{q_i}\gamma^\mu(1-\gamma_{5}) b)}(\bar{l}{\gamma_\mu }{\gamma_5}l).
\end{eqnarray}
Here $F_{\mu\nu}$ denotes the electromagnetic field strength tensor.
The operator that enters into the description of the $b\to q_i\Sigma\nu_l\bar{\nu_l}$ transitions is
\begin{equation}
     {\cal O}_{L}^{\nu_l}=\frac{e^2}{32\pi^2} \ {(\bar{q_i}\gamma^\mu(1-\gamma_{5}) b)} \ (\bar{\nu_l}{\gamma_\mu }(1-\gamma_5)\nu_l),
\end{equation}
 In SM, $C_R^{\nu_l}$ is negligible while $C_L^{\nu_l}=\frac{-X(x_t)}{sin^2\theta_W}$ with $x_t=m_t^2/m_W^2$ and ${X(x_t)}$ is an Inami-Lim function \cite{inami1981effects, buchalla1996weak}, which is given by 
 \begin{equation}
     X(x_t)=\frac{x}{8}\Big(\frac{2+x}{x-1}+\frac{3x-6}{(x-1)^2}lnx\Big).
 \end{equation}
With the effective Hamiltonians in the form (1-4), the structure of the free quark decay amplitudes are: 

\begin{eqnarray}
{\cal M}(b\to q_il\bar{l})=\frac{G_F}{\sqrt{2}}\frac{\alpha_{em} }{2\pi} V_{tb}V_{t{q_i}}^*\  &&[C_9^{eff} \ (\bar{q_i}\gamma^\mu(1-\gamma_5) b) \ (\bar{l}\gamma_\mu l)\nonumber\\ 
&&+ C_{10} \ (\bar{q_i}\gamma^\mu(1-\gamma_5) b) \ (\bar{l}\gamma_\mu \gamma_{5} l) \nonumber \\
&&-C_7^{eff} \ \frac{2m_b}{q^2}(\bar{q_i}i\sigma^{\mu\nu}q_\nu(1+\gamma_{5})b) \ (\bar{l}\gamma_\mu l)]
\end{eqnarray}
\begin{equation}
	{\cal M}(b\to q_i\Sigma\nu_l\bar{\nu}_l)=\frac{G_F}{\sqrt{2}}\frac{\alpha_{em} }{2\pi} V_{tb}V_{t{q_i}}^*\ C_L^{\nu_l} \ (\bar{q_i}\gamma^\mu(1-\gamma_{5}) b) \ (\bar{\nu}_l\gamma_\mu(1-\gamma_{5}) \nu_l)
\end{equation}
\subsection{Form factors and differential decay distribution}
For application of the quark level expressions in the description of rare semileptonic decays of $B_{(c)}$ mesons, it is necessary to calculate the matrix elements of the appropriate current operators (5-6) between the initial $|H_{in}(\vec{p}, S_{H_{in}})\rangle$ and final $|H_{f}(\vec{k}, S_{H_{f}})\rangle$ meson states ($H_{in}$ denotes either $B$ or $B_c$ and $H_f$ denotes the pseudoscalar mesons $K, D, D_s$). Such calculation requires an application of nonperturbative approaches, in which one needs to calculate the matrix elements: $\langle H_{f}(\vec{k}, S_{H_{f}})|\bar{q_i}\gamma^\mu(1-\gamma_5) b|H_{in}(\vec{p}, S_{H_{in}})\rangle$ and $\langle H_{f}(\vec{k}, S_{H_{f}})|\bar{q_i}i\sigma^{\mu\nu}q_\nu (1-\gamma_5)b|H_{in}(\vec{p}, S_{H_{in}})\rangle$. Here, the parts of the transition current containing $\gamma_5$ do not contribute. So we consider only the $\bar{q_i}\gamma^\mu b$ and also $ \bar{q_i}i\sigma^{\mu\nu}q_\nu b$ parts in the calculation of hadronic matrix elements. 
\par Considering Lorentz and parity invariance, the matrix elements of the weak currents for rare decays of meson $H_{in}$ to pseudoscalar meson $H_f$, can be parametrized by the invariant form factors ($f_+(q^2), f_-(q^2), f_T(q^2)$) as 
\begin{eqnarray}
J_V^\mu \equiv \langle H_f(k)|\bar{q_i}\gamma^\mu b|H_{in}(p)\rangle&=&f_+(q^2){\cal P}^\mu+f_-(q^2)q^\mu \\
J_T^\mu \equiv \langle H_{f}(k)|\bar{q_i}i\sigma^{\mu\nu}q_\nu b|H_{in}(p)\rangle&=&\frac{1}{M+m}\Big[{{\cal P}^\mu q^2-q^\mu{\cal P}q}\Big] \ f_T(q^2)
\end{eqnarray}
We specify our choice of momenta of participating particles as $p=k+k_1+k_2$, with $p^2=M^2$, $k^2=m^2$ and $k_1^2=k_2^2=\mu^2$, where $p, k$ represent the momenta of the initial and final meson; $k_1$ and $k_2$ are momenta of lepton $l^+$ and $l^-$.  
$M, m$ and $\mu$ are masses of initial meson $H_{in}$, final meson $H_f$ and lepton, respectively. We also specify our notation: $q=p-k$, ${\cal P}=p+k$ and the four-momentum transfer to the lepton pair $q=p-k$, which lies within the limit of $4\mu^2<q^2<(M-m)^2$.
\par For $B(B_c)\to K(D_{(s)})l\bar{l}$ - decays, the matrix element is written here as 
\begin{equation}
	{\cal M}=\frac{G_F}{\sqrt{2}}\frac{\alpha_{em}}{2\pi}V_{tb}V_{t{q_i}}^*[H_1^\mu(\bar{l}\gamma_\mu l)+H_2^\mu( \bar{l}\gamma_\mu\gamma_{5} l)]
\end{equation}
where the hadronic parts $H_1^{\mu}$ are obtained in terms of the invariant form factors and Wilson coefficients as
\begin{equation}
H_i^\mu=F^{(i)}_+{\cal P}^\mu+F^{(i)}_-q^\mu \ \ \ (i=1,2),    
\end{equation}
with
\begin{eqnarray}
	F^{(1)}_+&=&C_9^{eff}f_+(q^2)-C_7^{eff}f_T(q^2)\frac{2m_b}{M+m},\nonumber\\
	F^{(1)}_-&=&C_9^{eff}f_-(q^2)+C_7^{eff}f_T(q^2)\frac{2m_b}{M+m}\frac{{\cal P}q}{q^2},\nonumber\\
	F^{(2)}_\pm&=&C_{10}f_{\pm}(q^2).
\end{eqnarray}
From the invariant matrix element (9), the differential angular decay distribution for rare weak decays can be expressed in terms of hadronic $H_{ij}^{\mu\nu}$, $H^{\mu\nu}$ and leptonic tensors $L^{(k)}_{\mu\nu}$ in the form:
\begin{eqnarray}
&&H_{ij}^{\mu\nu}=H_i^\mu H_j^{\dagger\nu}, \ \   H^{\mu\nu}=H^\mu H^{\dagger\nu}. \nonumber \\
&&L_{\mu\nu}^{(1)}=k_{1\mu}k_{2\nu}+k_{2\mu}k_{1\nu},\ \ L_{\mu\nu}^{(2)}=g_{\mu\nu},\ \ L_{\mu\nu}^{(3)}=i\epsilon_{\mu\nu\alpha\beta}k_1^\alpha k_2^\beta. 
\end{eqnarray}
Then it is straightforward to find the differential angular decay distribution for $B(B_c)\to K(D_{(s)})l\bar{l}$ in the form \cite{faessler2002exclusive, ebert2010rare, choi2010light}
\begin{eqnarray}
	\frac{d^2\Gamma}{dq^2dcos\theta}&=&\frac{G_F^2}{(2\pi)^3}\Big(\frac{\alpha_{em}|V_{tb}V_{t{q_i}}^*|}{2\pi}\Big)^2\frac{|\vec{k}|}{8M^2} \sqrt{\Big(1-\frac{4\mu^2}{q^2}\Big)}\frac{1}{2}\Big\{L_{\mu\nu}^{(1)}(H_{11}^{\mu\nu}+H_{22}^{\mu\nu})\nonumber \\&-&\frac{1}{2}L_{\mu\nu}^{(2)}(q^2H_{11}^{\mu\nu}+(q^2-4\mu^2)H_{22}^{\mu\nu})+L_{\mu\nu}^{(3)}(H_{12}^{\mu\nu}+H_{21}^{\mu\nu})\Big\}.
\end{eqnarray}
\par For $B(B_c)\to K(D_{(s)})\Sigma\nu_l\bar{\nu_l}$ - decays, the invariant matrix element is
\begin{equation}
	{\cal M}=\frac{G_F}{\sqrt{2}}\frac{\alpha_{em}}{2\pi}V_{tb}V_{t{q_i}}^* \ C_L^{\nu_l} \ [H^\mu \ \bar{\nu_l}\gamma_\mu(1-\gamma_{5})\nu_l],\nonumber
\end{equation}
Note that for such decay mode, the form factor $f_T(q^2)$ does not contribute, since it is related to the photon vertex ($\sigma^{\mu\nu}q_\nu$).
Therefore, the hadronic part $H^\mu$ is obtained here in the simple form:
\begin{equation}
H^{\mu}=f_+(q^2){\cal P}^\mu+f_-(q^2)q^\mu.    
\end{equation} 
Then the differential angular decay distribution for $B(B_c)\to K(D_{(s)})\Sigma\nu_l\bar{\nu_l}$ is obtained in the form \cite{ebert2010rare, choi2010light}
\begin{equation}
 \frac{d^2\Gamma}{dq^2dcos\theta}=\frac{G_F^2}{(2\pi)^3}\Big(\frac{\alpha_{em}|V_{tb}V_{t{q_i}}^*|}{2\pi}\Big)^2\frac{3|\vec{k}|}{8M^2} \ |C_L^{\nu_l}|^2\Big\{L_{\mu\nu}^{(1)}-\frac{q^2}{2}L_{\mu\nu}^{(2)}+L_{\mu\nu}^{(3)}\Big\}H^{\mu\nu},
\end{equation}
where factor 3 comes due to the contribution from all three neutrino modes. 
\subsection{Helicity amplitude and differential decay rate}
It is well known that physical observable such as hadron tensor is conveniently expressed on the helicity basis \cite{faessler2002exclusive, nayak2022exclusive} in which, helicity form factors are obtained in terms of the Lorentz invariant form factors. While doing the Lorentz contraction in Eq.(13, 15) using the helicity amplitudes, one considers four covariant helicity projections $\epsilon^\mu(m)$ with $m = +, -, 0$ and $m = t$. In our attempt to study the lepton mass effects in the rare semileptonic decay modes, we need to consider the time component of the polarization $\epsilon^\mu(m=t)$ in addition to its other three components with $m = +, -, 0$. 
\par Using orthogonality and the completeness relations satisfied by helicity projections, the Lorentz contraction in Eq. (13, 15) can be written as
\begin{eqnarray}
 	L^{\mu\nu(k)}H_{\mu\nu}^{ij}&&=L^{(k)}(m, n)g_{mm^{'}}g_{nn^{'}}H^{ij}(m^{'}, n^{'})\nonumber\\ 
  L^{\mu\nu(k)}H_{\mu\nu}&&=L^{(k)}(m, n)g_{mm^{'}}g_{nn^{'}}H(m^{'}, n^{'})
\end{eqnarray}
with $g_{mn}$ = dia(+, -, -, -). Here the lepton and hadron tensors are introduced in the space of helicity components as:
\begin{eqnarray}
L^{(k)}(m, n)&=&\epsilon^\mu(m) \epsilon^{\dagger\nu}(n)L_{\mu\nu}^{(k)}, \nonumber\\
H^{ij}(m^{'}, n^{'})&=&\epsilon^\mu(m^{'}) \epsilon^{\dagger\nu}(n^{'})H^{ij}_{\mu\nu}, \nonumber \\
H(m^{'}, n^{'})&=&\epsilon^\mu(m^{'}) \epsilon^{\dagger\nu}(n^{'})H_{\mu\nu}.    
\end{eqnarray}
\par For the sake of convenience, we consider here two frames of reference: (i) the $l\bar{l}$ and $\Sigma\nu_l\bar{\nu_l}$ centre-of-mass frame and (ii) the parent $B_{(c)}$-meson rest frame. We evaluate lepton tensor $L^{(k)}(m, n)$ in the $l\bar{l}$ and $\Sigma\nu_l\bar{\nu_l}$ centre-of-mass frame and hadron tensor $H^{ij}(m^{'}, n^{'})$ and $H(m^{'}, n^{'})$ in the $B_{(c)}$- rest frame. In the $B_{(c)}$- rest frame, the four momenta: ($p^{\mu}, k^{\mu}, q^{\mu}$) and polarization vectors: ($\epsilon^\mu(\pm)$, $\epsilon^{\mu}(0)$ and $\epsilon^{\mu}(t)$) are specified as $p^\mu=(M, 0, 0, 0)$, $k^\mu=(E_k, 0, 0, -|\vec{k}|)$, $q^\mu=(q^0, 0, 0, |\vec{k}|)$ and $\epsilon^\mu(t)=\frac{1}{\sqrt{q^2}}(q^0, 0, 0, |\vec{k}|)$, $\epsilon^\mu(\pm)=\frac{1}{\sqrt{2}}(0, \mp1, -i, 0)$,
$\epsilon^\mu(0)=\frac{1}{\sqrt{q^2}}(|\vec{k}|, 0, 0, q^0)$, where $E_k=\frac{M^2+m^2-q^2}{2M}$,  $q^0=\frac{M^2-m^2+q^2}{2M}$. The hadronic amplitude $H_i^\mu$ is then cast in the helicity frame. 
\par For $B_{(c)}\to K(D_{(s)})l\bar{l}$ - decays, the helicity form factors $H^i(m)$ are obtained in terms of invariant form factor as:
\begin{eqnarray}
H^i(t)&=&\frac{1}{\sqrt{q^2}} \ ({\cal P}qF_+^{(i)}+q^2F_-^{(i)}),\nonumber \\
H^i(\pm)&=&0, \nonumber \\
H^i(0)&=&\frac{2M|\vec{k}|}{\sqrt{q^2}} \ F_+^{(i)}.
\end{eqnarray}
Similarly, for $B(B_c)\to K(D_{(s)})\Sigma\nu_l\bar{\nu_l}$ - decays, the helicity form factors $H(m)$ are obtained in the form:
\begin{eqnarray}
H(t)&=&\frac{1}{\sqrt{q^2}}({\cal P}qf_+(q^2)+q^2f_-(q^2)),\nonumber \\
H(\pm)&=&0, \nonumber \\
H(0)&=&\frac{2M|\vec{k}|}{\sqrt{q^2}}f_+(q^2).
\end{eqnarray}
Using the space and time components of the four momenta: ($q^\mu, k^\mu, q^\mu$) and helicity projections in the $l\bar{l}$ and $\Sigma\nu_l\bar{\nu_l}$- centre-of-mass frame: 
\begin{eqnarray}
q^\mu&=&(\sqrt{q^2}, 0, 0, 0),\nonumber \\
k_1^\mu&=&(E_l, |\vec{k}_1|sin\theta cos\chi, -|\vec{k}_1|sin\theta sin\chi, |\vec{k}_1|cos\theta), \nonumber\\
k_2^\mu&=&(E_l, -|\vec{k}_1|sin\theta cos\chi, -|\vec{k}_1|sin\theta sin\chi, -|\vec{k}_1|cos\theta),
\end{eqnarray}
 with $E_l=\sqrt{q^2}/2$ and $|\vec{k}_1|=\sqrt{q^2-4\mu^2}/2$ for two lepton in final states and $E_l=|\vec{k}_1|=\sqrt{q^2}/2$ for two neutrinos in final states, the longitudinal and time components of the polarization vector are reduced to $\epsilon^\mu(0)=(0, 0, 0, 1)$, $\epsilon^\mu(t)=(1, 0, 0, 0)$, whereas its transverse part $\epsilon^\mu(\pm)$ remains unchanged. Using these factors, it is straightforward to calculate the helicity representation $L^{(k)}(m, n)$ of the lepton tensor (17).
 \par In the present analysis, we do not consider the azimuthal $\chi$
distribution of the lepton pair and therefore integrate over the azimuthal angle dependence of the lepton tensor leading to the differential ($q^2, cos\theta$) distribution of the rare semileptonic decays of $B(B_c)$ mesons. \\
 (a) For $B(B_c)\to K(D_{(s)})l\bar{l}$, the differential ($q^2, cos\theta$) distribution is obtained in form:
\begin{eqnarray}
	\frac{d^2\Gamma}{dq^2d(cos\theta)}&=&\frac{3}{8}(1+cos^2\theta).\frac{1}{2}\Big(\frac{d\Gamma_U^{11}}{dq^2}+\frac{d\Gamma_U^{22}}{dq^2}\Big)+\frac{3}{4}sin^2\theta.\frac{1}{2}\Big(\frac{d\Gamma_L^{11}}{dq^2}+\frac{d\Gamma_L^{22}}{dq^2}\Big)\nonumber\\
	&+&\sqrt{1-\frac{4\mu^2}{q^2}}.\frac{3}{4}cos\theta.\frac{d\Gamma_P^{12}}{dq^2}+\frac{3}{4}sin^2\theta.\frac{1}{2}\frac{d\tilde{\Gamma}_U^{11}}{dq^2}-\frac{3}{8}(1+cos^2\theta).\frac{d\tilde{\Gamma}_U^{22}}{dq^2}\nonumber\\
	&+&\frac{3}{2}cos^2\theta.\frac{1}{2}\frac{d\tilde{\Gamma}_L^{11}}{dq^2}-\frac{3}{4}sin^2\theta.\frac{d\tilde{\Gamma}_L^{22}}{dq^2}+\frac{1}{4}\frac{d\tilde{\Gamma}_S^{22}}{dq^2}.
\end{eqnarray}
Then integrating over $cos\theta$, the differential decay width is obtained as
\begin{eqnarray}
	\frac{d\Gamma}{dq^2}&=&\frac{1}{2}\Big(\frac{d\Gamma_U^{11}}{dq^2}+\frac{d\Gamma_U^{22}}{dq^2}+\frac{d\Gamma_L^{11}}{dq^2}+\frac{d\Gamma_L^{22}}{dq^2}\Big)\nonumber\\
	&+&\frac{1}{2}\frac{d\tilde{\Gamma}_U^{11}}{dq^2}-\frac{d\tilde{\Gamma}_U^{22}}{dq^2}
	+\frac{1}{2}\frac{d\tilde{\Gamma}_L^{11}}{dq^2}-\frac{d\tilde{\Gamma}_L^{22}}{dq^2}+\frac{1}{2}\frac{d\tilde{\Gamma}_S^{22}}{dq^2},
\end{eqnarray}
Out of nine terms in the r.h.s of Eq.(22), five terms identified by tilde rates $\tilde{\Gamma}_i$ are liked with the lepton mass and other terms identified by $\Gamma_i$ are lepton mass independent. Both are related via a flip factor $\frac{2\mu^2}{q^2}$ as
\begin{equation}
    \frac{d\tilde{\Gamma}_X^{ij}}{dq^2}=\frac{2\mu^2}{q^2} \ \frac{d\Gamma_X^{ij}}{dq^2}.
\end{equation}
The tilde rates do not contribute to the decay rate in the vanishing lepton mass limit. They have minimal contributions for rare weak decays to $e$ and $\mu$-modes whereas their contribution to decays to $\tau$-mode is significant. Therefore, the tilde modes are crucial in the evaluation of the lepton mass effect in the rare semileptonic decay modes.\\
The differential partial helicity rates $d{\Gamma}_X^{ij}/{dq^2}$ are obtained in the form
\begin{equation}
	\frac{d\Gamma_X^{ij}}{dq^2}=\frac{G_F^2}{(2\pi)^3} \ \Big(\frac{\alpha_{em}|V_{tb}V_{t{q_i}}^*|}{2\pi}\Big)^2\frac{|\vec{k}|q^2}{12M^2} \ \sqrt{1-\frac{4\mu^2}{q^2}} \ H_X^{ij}.	
\end{equation}
Here $H_X^{ij}$ $(X=U, L, P, S; i, j=1, 2)$ represents a standard set of helicity structure functions given by linear combinations of helicity components of hadron tensor $H^{ij}(m^{'}, n^{'})= H^i(m^{'}) H^{\dagger j}(n^{'})$ and $H(m, n)= H(m) H^{\dagger }(n)$: 
\begin{eqnarray}
	H_U^{ij}&=&Re(H_+^iH_+^{\dagger j})+Re(H_-^iH_-^{\dagger j}): Unpolarized-transverse,\nonumber\\
	H_P^{ij}&=&Re(H_+^iH_+^{\dagger j})-Re(H_-^iH_-^{\dagger j}): Parity-odd,\nonumber\\
	H_L^{ij}&=&Re(H_0^iH_0^{\dagger j}): Longitudinal,\nonumber\\
	H_S^{ij}&=&3Re(H_t^iH_t^{\dagger j}): Scalar.
\end{eqnarray}
(b) For $B(B_c)\to K(D_{(s)})\Sigma\nu_l\bar{\nu_l}$, the differential $(q^2, cos\theta)$ distribution is obtained as:
\begin{equation}
\frac{d^2\Gamma}{dq^2d(cos\theta)}=\frac{G_F^2}{(2\pi)^3}\Big(\frac{\alpha_{em}|V_{tb}V_{t{q_i}}^*|}{2\pi}\Big)^2 \ \frac{3|\vec{k}|q^2}{16M^2} \ C_L^{\nu_l} \ sin^2\theta \ |H(0)|^2
\end{equation}
Integrating over $cos\theta$, the differential decay width is reduced to the form:
\begin{equation}
\frac{d\Gamma}{dq^2}=\frac{G_F^2}{(2\pi)^3}\Big(\frac{\alpha_{em}|V_{tb}V_{t{q_i}}^*|}{2\pi}\Big)^2 \ \frac{|\vec{k}|q^2}{4M^2} \ |C_L^{\nu_l}|^2 |H(0)|^2
\end{equation}
\subsection{Longitudinal lepton polarization asymmetry}
An interesting observable, the Lepton Polarization Asymmetry (LPA), is defined as \\
\begin{equation}
 P_L(q^2)=\frac{d\Gamma_{h=-1}/dq^2-d\Gamma_{h=+1}/dq^2}{d\Gamma_{h=-1}/dq^2+d\Gamma_{h=+1}/dq^2}   
\end{equation}
where $h=+1(-1)$ denotes right (left) handed $l^-$ in the final state. From the differential decay width expressions for $h=+1(-1)$, it is straightforward to obtain $P_L(q^2)$ in terms of the invariant form factors and Wilson coefficients as:
\begin{equation}
P_L(q^2)=\frac{2\sqrt{1-\frac{4\mu^2}{q^2}}|\vec k|^2C_{10}f_+(q^2)\Big[f_+(q^2)ReC_9^{eff}-\frac{2m_b}{M+m}C_7^{eff }f_T(q^2)\Big]}{|\vec k|^2(1+\frac{2\mu^2}{q^2}){\cal F}_1+3\mu^2{\cal F}_2}
\end{equation}
where,
\begin{eqnarray}
{\cal F}_1&&=\Big|f_+(q^2)C_9^{eff}-\frac{2m_b}{M+m}C_7^{eff }f_T(q^2)\Big|^2+|C_{10}f_+(q^2)|^2 \nonumber \\ 
{\cal F}_2&&=|C_{10}|^2\Big[(1+\frac{m^2}{M^2}-\frac{q^2}{2M^2})|f_+(q^2)|+(1-\frac{m^2}{M^2})f_+(q^2)f_-(q^2)+\frac{q^2}{2M^2}|f_-(q^2)|^2\Big]\nonumber
\end{eqnarray}
Due to the experimental difficulties in the study of the polarizations of each lepton, which depends on $q^2$ and the Wilson coefficients, it is better to eliminate the  $q^2$-dependence of the LPA by considering its averaged form over the entire kinematical range. The averaged LPA is defined by
\begin{equation}
    \langle P_L\rangle=\frac{\int_{4\mu^2}^{(M-m)^2}\ P_L(q^2) \ \frac{dBF}{dq^2} \ dq^2}{\int_{4\mu^2}^{(M-m)^2} \ \frac{dBF}{dq^2} \ dq^2}
\end{equation}
\section{Invariant transition form factors in the relativistic independent quark (RIQ) model}
The decay process physically occurs in the momentum eigenstate of the participating mesons. Therefore, in the field-theoretic description of the decay process, the participating meson-bound states are represented here by the appropriate momentum wave packets with suitable momentum and spin distribution between the constituent quark-antiquark pair in the corresponding meson core. In the relativistic independent particle picture of the RIQ model, the constituent quark and antiquark are assumed to be in the state of independent and relativistic motion inside the meson-bound state: $|M(\vec{p},S_{M})\rangle$, for example, with constituent quark and antiquark momentum $q_1$ and $\bar{q_2}$, respectively. The momentum probability amplitude of the constituent quark and antiquark in the participating meson ground state, are extracted here from the momentum projection of the static quark orbitals, obtainable in the present model framework. The corresponding expressions for the momentum probability amplitude ${\cal G}_{q_1}(\vec{p}_{q_1})$ for the constituent quark $q_1$ and $\tilde{{\cal G}}_{q_2}(\vec{p}_{q_2})$ for the antiquark $\bar{q_2}$ are given in (A6) of the Appendix. In this model, we construct an effective momentum distribution function ${\cal G}_M(\vec{p}_{q_1},\  \vec{p}_{q_2})$ in the form ${\cal G}_M(\vec{p}_{q_1},\  \vec{p}_{q_2})=\sqrt{G_{q_1}(\vec{p}_{q_1}) \tilde{G}_{q_2}(\vec{p}_{q_2})}$ in a straightforward extension of ansatz of Margolis and Mendel \cite{margolis1983bag} in their bag model description of the meson bound-state. With the effective momentum distribution function ${\cal G}_M(\vec{p}_{q_1},\  \vec{p}_{q_2})$, representing the bound-state character, the meson bound state in the RIQ model is taken in the form
\begin{eqnarray}
  |M(\vec{p},S_{M})\rangle&&={\hat{\Lambda}}(\vec{p},S_{M})|(\vec{p}_{q_1},\lambda_{q_1});(\vec{p}_{q_2},\lambda_{q_2})\rangle\nonumber\\&&={\hat{\Lambda}}(\vec{p},S_{M})\hat{b}^\dagger_{q_1}(\vec{p}_{q_1},\lambda_{q_1}) \hat{\tilde{b}}^\dagger_{q_2}(\vec{p}_{q_2},\lambda_{q_2})|0\rangle,
  \end{eqnarray}
where, $|(\vec{p}_{q_1},\lambda_{q_1});(\vec{p}_{q_2},\lambda_{q_2})\rangle$ is the Fock-space representation of the unbound quark and antiquark in a color-singlet configuration with their respective momentum and spin: $(\vec{p}_{q_1},\lambda_{q_1})$ and $(\vec{p}_{q_2},\lambda_{q_2})$. Here $\hat{b}^\dagger_{q_1}(\vec{p}_{q_1},\lambda_{q_1})$ and $\hat{\tilde{b}}^\dagger_{q_2}(\vec{p}_{q_2},\lambda_{q_2})$ are the quark and antiquark creation operators and ${\hat{\Lambda}}(\vec{p},S_{M})$ is a bag-like operator taken here in the integral form:
\begin{equation}
{\hat{\Lambda}}(\vec{p},S_{M})=\frac{\sqrt{3}}{\sqrt{N_{M}(\vec{p})}}\sum_{\lambda_{q_1},\lambda_{q_2}}\zeta_{q_1,{q_2}}^{M}\int d\vec{p}_{q_1}\ d\vec{p}_{q_2} \delta^{(3)}(\vec{p}_{q_1}+\vec{p}_{q_2}-\vec{p})\ {\cal G}_{M}(\vec{p}_{q_1}, \vec{p}_{q_2}),       
\end{equation}
where $\sqrt{3}$ is the effective color factor, $\zeta_{q_1,{q_2}}^{M}$ is the $SU(6)$ spin-flavor coefficients for the meson state $|{M}(\vec p, S_{M})\rangle$. Finally, $N_{M}(\vec{p})$ represents the meson state normalization, which is obtained in the integral form:
\begin{equation}
 N_{M}(\vec{p})=\frac{1}{(2\pi)^3{2E_p}}\int d\vec{p}_{q_1} |{\cal G}_{M}(\vec{p}_{q_1}, \vec p_{q_2})|^2=\frac{{\bar{N}_{M}(\vec{p})}}{(2\pi)^3{2E_p}}. \end{equation} 
 by imposing a normalization condition $\langle M(\vec{p}^{'})|M(\vec{p})\rangle={(2 \pi)}^3 2E_p \delta^{(3)}(\vec{p}-\vec{p}^{'})$. A brief description of the quark orbitals obtained in the Dirac framework using the RIQ model conventions and corresponding momentum probability amplitudes. $(G_{q_1}(\vec{p}_{q_1}), \tilde{G}_{q_2}(\vec{p}_{q_2}))$ for the constituent quark and antiquark are given in the Appendix.
\par Now using the wavepacket representation of the initial $|H_{in}\rangle$ and final $|H_f\rangle$ meson states, the invariant hadronic matrix elements $J_V^\mu$ and $J_T^\mu$ (7, 8) are calculated in the parent meson rest frame, yielding 
\begin{eqnarray}
J_V^\mu \equiv\langle H_f(k)|\bar{q_i}\gamma^\mu b|H_{in}(0)\rangle &&=\sqrt{\frac{4E_kM}{\bar{N}_{H_{in}}(0)\bar{N}_{H_f}(\vec{k})}}\int \frac{d\vec{p}_b}{\sqrt{2E_{p_b} \ 2E_{p_b+k}}} \nonumber \\ 
&&{\cal G}_{H_{in}}(\vec{p}_b, -\vec{p}_b){\cal G}_{H_f}(\vec{p}_b+\vec{k}, -\vec{p}_b) \langle S_{H_f}|J_V^\mu(0)|S_{H_{in}}\rangle, 
\end{eqnarray}
\begin{eqnarray}
J_T^\mu \equiv\langle H_f(k)|\bar{q_i}i\sigma^{\mu\nu}{q_\nu }b|H_{in}(0)\rangle &&=\sqrt{\frac{4E_kM}{\bar{N}_{H_{in}}(0)\bar{N}_{H_f}(\vec{k})}}\int \frac{d\vec{p}_b}{\sqrt{2E_{p_b} \ 2E_{p_b+k}}} \nonumber \\ 
&&{\cal G}_{H_{in}}(\vec{p}_b, -\vec{p}_b){\cal G}_{H_f}(\vec{p}_b+\vec{k}, -\vec{p}_b) \langle S_{H_f}|J_T^\mu(0)|S_{H_{in}}\rangle, 
\end{eqnarray}
where $E_{p_b}$ and $E_{p_b+k}$ stand for the energy of the non-spectator quark of the parent and daughter meson, respectively. $\langle S_{H_f}|J_V^\mu(0)|S_{H_{in}}\rangle$ and $\langle S_{H_f}|J_T^\mu(0)|S_{H_{in}}\rangle$ represent symbolically the spin matrix elements of vector and tensor current, respectively.
\par For $0^-\to 0^-$ transitions, the axial vector current does not
contribute. The spin matrix elements corresponding to the
non-vanishing vector and tensor currents are obtained in the form:
\begin{eqnarray}
  \langle S_{H_f}|V^0|S_{H_{in}}\rangle&=&\frac{(E_{p_b}+m_b)(E_{p_b+k}+m_{d/s})+|\vec{p_b}|^2}{\sqrt{(E_{p_b}+m_b)(E_{p_b+k}+m_{d/s})}} \\
   \langle S_{H_f}|V^i|S_{H_{in}}\rangle&=&\frac{(E_{p_b}+m_b) \ k^i}{\sqrt{(E_{p_b}+m_b)(E_{p_b+k}+m_{d/s})}}\\
    \langle S_{H_f}|T^0|S_{H_{in}}\rangle&=&-\frac{(E_{p_b}+m_b) \ |\vec{k}|^2}{\sqrt{(E_{p_b}+m_b)(E_{p_b+k}+m_{d/s})}}
\end{eqnarray}
    
With the above spin matrix elements, the expressions for hadronic amplitudes (34, 35) are compared with corresponding expressions from the Eqs. (7, 8) yielding the model expression of form factors $f_{\pm}(q^2)$ and $f_T(q^2)$ in the form
\begin{eqnarray}
f_{\pm}(q^2)&&=\frac{1}{2M}\sqrt{\frac{ME_k}{\bar{N}_{H_{in}}(0)\bar{N}_{H_f}(\vec{k})}}\int d\vec{p}_b \  {\cal G}_{H_{in}}(\vec{p}_b, -\vec{p}_b) \  {\cal G}_{H_f}(\vec{p}_b+\vec{k}, -\vec{p}_b) \nonumber \\
&&\times\Big[\frac{(E_{p_b}+m_b)(E_{p_b+k}+m_{d/s})\pm(M\mp E_k)(E_{p_b}+m_b)+|\vec{p}_b|^2}{\sqrt{E_{p_b} \ E_{p_b+k} \ (E_{p_b}+m_b) \ (E_{p_b+k}+m_{d/s})}}\Big]
\end{eqnarray}
\begin{eqnarray}
f_T(q^2)=-\frac{(M+m)}{2M}\sqrt{\frac{ME_k}{\bar{N}_{{H_{in}}}(0)\bar{N}_{H_f}(\vec{k})}}&& \int d\vec{p}_b \  {\cal G}_{H_{in}}(\vec{p}_b, -\vec{p}_b) \  {\cal G}_{H_f}(\vec{p}_b+\vec{k}, -\vec{p}_b)\nonumber \\
&&\times\sqrt{\frac{(E_{p_b}+m_b)}{E_{p_b} \ E_{p_b+k} \ (E_{p_b+k}+m_{d/s})}}
\end{eqnarray}
\section{Numerical results and discussion}
In this section, we present our numerical results and discussion on exclusive rare semileptonic decays: $B(B_c)\to K(D_{(s)})l\bar{l}$ and $B(B_c)\to K(D_{(s)})\Sigma\nu_l\bar{\nu_l}$. The input parameters used in the numerical calculation: the quark masses ($m_u=m_d$, $m_s$, $m_c$, $m_b$) and potential parameters ($a$, $V_0$) have been fixed in this model by fitting our predicted meson masses with the data of heavy and heavy-light flavored mesons in their ground state \cite{dash2024purely}. The input parameters, so fixed from hadron spectroscopy, are shown in Table I. 
\begin{table}[!hbt]
\renewcommand{\arraystretch}{0.1}
	\centering
	\setlength\tabcolsep{2pt}
	\caption{The quark masses $m_q$ (in $GeV$) and potential parameters:  $a$ (in $GeV^3$) and $V_0$ (in $GeV$)} 

	\label{tab1}
	
		\begin{tabular}{cccccc}
			\hline
   &&&&&\\
 \ \ $m_u=m_d$& \ \ \ $m_s$& \ \ \ $m_c$& \ \ \ $m_b$& \ \ \ $a$ \ \ &$\ \ V_0$ \ \ \\
&&&&&\\
 \ \ 0.26& \ \ \ 0.49& \ \ \ 1.64& \ \ \ 4.92&\ \ \ 0.023& \ \ -0.307\\
 &&&&&\\
 \hline
\end{tabular}
\end{table}
\begin{table}[!hbt]
 	\renewcommand{\arraystretch}{1.25}
 	\centering
 	\setlength\tabcolsep{12pt}
 	\caption{Numerical Inputs}
 	\label{tab2}
 	\begin{tabular}{cl}	
  \hline
  Parameter& \ \ \ \ \ \ Value \cite{navas2024review}\\
\hline
$G_F$&$1.1663788\times 10^{-5}$ GeV$^{-2}$\\
$M_{B_c^{+ }}$& $6274.47\pm0.27\pm0.17$ MeV\\
$M_{B^{+ }}$& $5279.42\pm0.08$ MeV \\
$m_{D^{+ }}$& $1869.5\pm0.4$ MeV \\
$m_{D_s^{+ }}$& $1969.0\pm1.4$ MeV \\
$m_{K^{+ }}$& $493.677\pm0.005$ MeV\\
$\tau_{B_c^{+ }}$& $0.51\pm0.009$ ps \\
$\tau_{B^{ +}}$& $1.638\pm0.004$ ps \\
$|V_{td}|$&$0.00858^{+0.00019}_{-0.00017}$ \\	
$|V_{ts}|$&$0.04111^{+0.00077}_{-0.00068}$\\		
$|V_{tb}|$&$0.999118^{+0.000029}_{-0.000034}$\\
\hline
 \end{tabular}
 \end{table}
\par The phenomenological input parameters: the Fermi Coupling constant $G_F$, participating meson masses, lifetime $\tau$ of decaying $ B_{(c)}$ mesons and the precise global fit values of CKM parameters, taken from PDG \cite{navas2024review} are listed in Table II. Although our predictions \cite{dash2024purely} of ground state heavy meson masses are overall in good agreement with the corresponding experimental data, we used the observed masses of ($B_{(c)}^+$, $D_{(s)}^+$ and $K^+$) mesons \cite{navas2024review} in our calculation to reduce possible theoretical uncertainties. We use the values of the Wilson Coefficients for decays to two lepton final states, $C_7^{eff}=-0.313$, $C_9^{eff}=4.344$ and $C_{10}=-4.669$ from Refs. \cite{buras1995effective, azizi2008analysis}. For decays to two neutrino final states, the value of the effective Wilson Coefficient $C_L^{\nu_l}=-6.32(7)$ used in the present calculation, is taken from Refs. \cite{buras2015b, altmannshofer2009new, buchalla1993qcd, buchalla1999rare, allwicher2024understanding, misiak1999qcd, brod2011two}. 
\begin{table}[!hbt]
	\renewcommand{\arraystretch}{0.75}
	\centering
	\setlength\tabcolsep{2pt}
	\caption{Results for form factors at $q^2\to q^2_{min}$ of $B({B_{c}})\to K(D_{(s)})\mu^+\mu^-/\Sigma\nu_l\bar{\nu_l}$ decays}
	\label{tab3}
\begin{tabular}{lcccc}
\hline Decay mode \ \ &&\ \ $f_+(q^2)$&\ \ $f_-(q^2)$&\ \ $f_T(q^2)$ \\
		\hline
            &This work&0.044&-0.067&-0.072\\
            ${B_c}\to D\mu^+\mu^-/ \Sigma\nu_l{\bar{\nu_l}}$&\cite{ebert2010rare}&0.081&0.081&0.061\\
            &\cite{choi2010light}&0.079&-0.070&-0.108\\
    \hline
		&This work&0.078&-0.107&-0.122\\
  ${B_c}\to D_s \mu^+\mu^-/ \Sigma\nu_l{\bar{\nu_l}}$&\cite{ebert2010rare}&0.129&0.129&0.098\\
  &\cite{choi2010light}&0.126&-0.099&-0.168\\
  \hline
		&This work&0.166&-0.173&-0.186\\
  $B\to K \mu^+\mu^-/ \Sigma\nu_l{\bar{\nu_l}}$&\cite{ebert2010rare}&0.242&0.242&0.258\\
	
  \hline
\end{tabular}
\end{table}
 \begin{table}[!hbt]
	\renewcommand{\arraystretch}{0.75}
	\centering
	\setlength\tabcolsep{2pt}
	\caption{Results for form factors at $q^2\to q^2_{max}$ of $B({B_{c}})\to K(D_{(s)})\mu^+\mu^-/\Sigma\nu_l\bar{\nu_l}$ decays}
	\label{tab4}
\begin{tabular}{lcccc}
\hline Decay mode \ \ &&\ \ $f_+(q^2)$&\ \ $f_-(q^2)$&\ \ $f_T(q^2)$ \\
		
    \hline
  	&This work&2.755 &-3.807 &-4.258\\
  ${B_c}\to D\mu^+\mu^-/\Sigma\nu_l\bar{\nu_l}$&\cite{choi2010light}&0.754&-0.721&-1.012\\  
  \hline
		&This work&3.924&-2.749&-3.250\\
  $B_c \to D_s\mu^+\mu^-/\Sigma\nu_l\bar{\nu_l}$&\cite{ebert2010rare}&2.662&0.669&2.151\\
		&\cite{choi2010light}&0.874&-0.722&-1.155\\

  \hline
		&This work&1.408&-1.074&-1.357\\
 $B\to K \mu^+\mu^-/\Sigma\nu_l\bar{\nu_l}$&\cite{ebert2010rare}&2.991&0.6601&0.794\\
		
  \hline
\end{tabular}
\end{table}
 \par With the input parameters (Table-I), the Lorentz invariant form factors: ($f_+, f_-, f_T$) representing decay amplitudes are calculated from the overlapping integral of participating meson wave functions. Before evaluating numerically the physical quantities of interest, we study the $q^2$- dependence of the invariant form factors in the accessible kinematical range. Since one of our objectives here is to study the Lepton mass effect in the rare semileptonic decays, we plot the $q^2$- dependence of the form factors in $B_{(c)}$-meson decays to their $e$, $\mu$ and $\tau$- modes. Our results are shown in Fig.1 and Fig.2, where we find that the behaviour of form factors in $e$ and $\mu$- mode overlap in the entire kinematic range: $(q^2_{min}\simeq0) \leq q^2 \leq q^2_{max}$. This is because of an insignificant change in the phase space boundary going from $e$ to $\mu$- mode. As one can see here, the maximal lepton energy shift $\frac{{m_\mu}^2-{m_e^2}}{2M}$ is invisible at the usual scale of the plot. On the other hand, for the decays in the $\tau$- mode, the relevant form factors behave differently throughout the accessible kinematical range of $q^2_{min} \leq q^2 \leq q^2_{max}$, where $q^2_{min}$ is +ve and away from $q^2\to 0$. The $\tau$- phase space, as compared to those in decays to $e$ and $\mu$-modes, is considerably reduced and shifted to a large $q^2$ region. In the present study we, therefore, consider decays to their $\mu^+\mu^-$ and $\tau^+\tau^-$ modes only for evaluating the lepton mass effect on the rare semileptonic $B_{(c)}$ meson decays
 \begin{figure}[!hbt]
	\centering
	\includegraphics[width=0.6\textwidth]{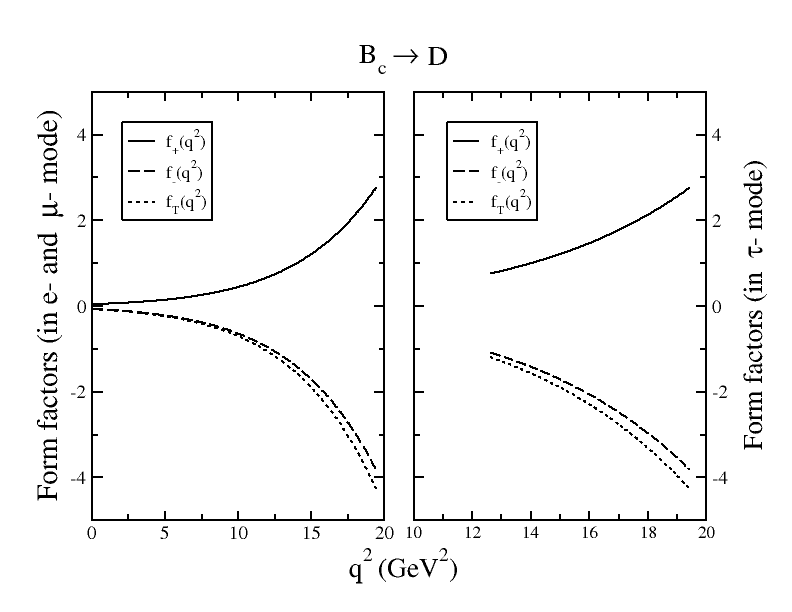}
	\includegraphics[width=0.6\textwidth]{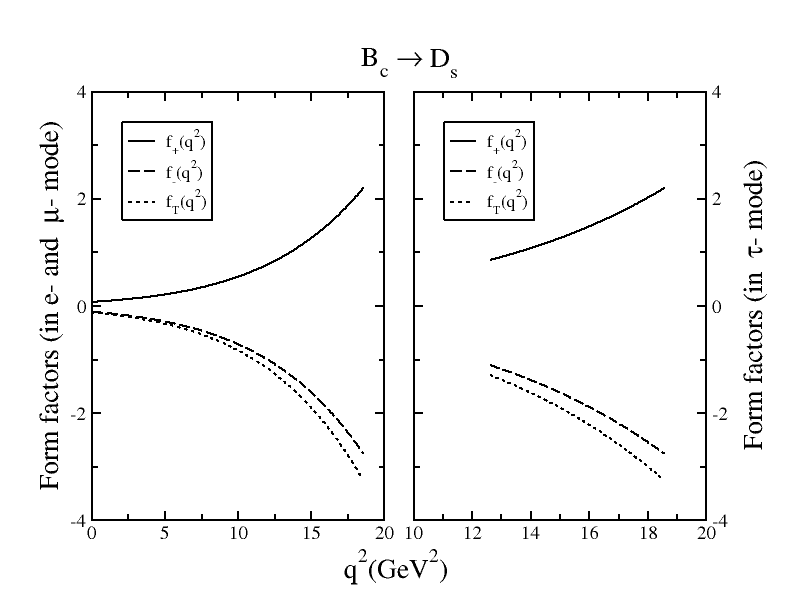}
	\includegraphics[width=0.6\textwidth]{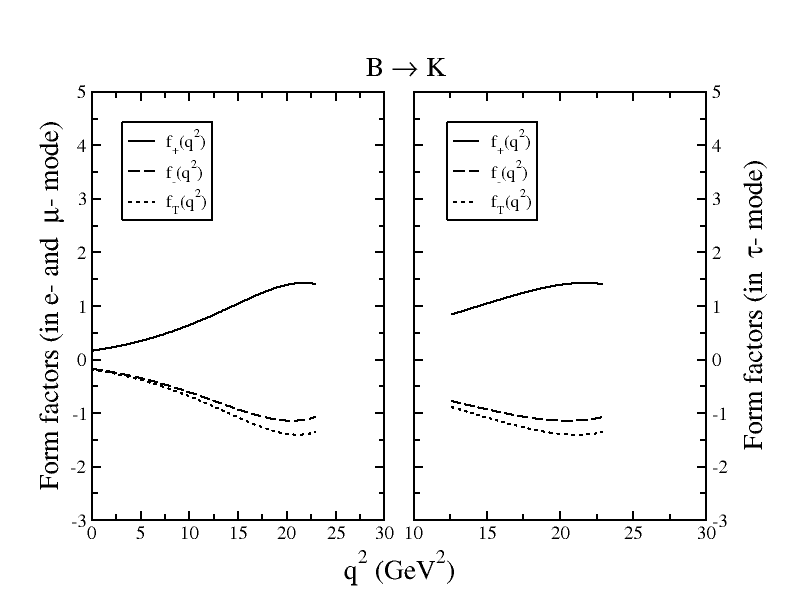}
	\caption{$q^2$-dependence of the form factors.}
\end{figure}
\begin{figure}[!hbt]
	\centering
	\includegraphics[width=0.6\textwidth]{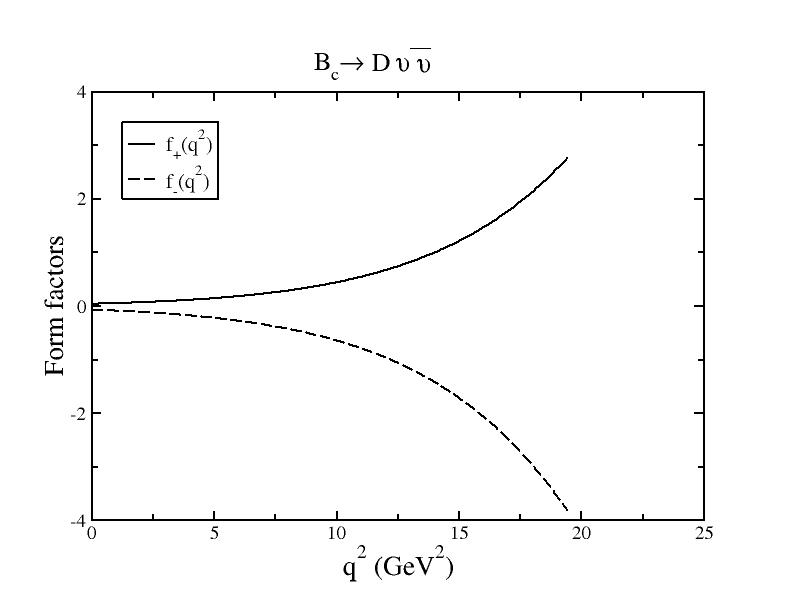}
	\includegraphics[width=0.6\textwidth]{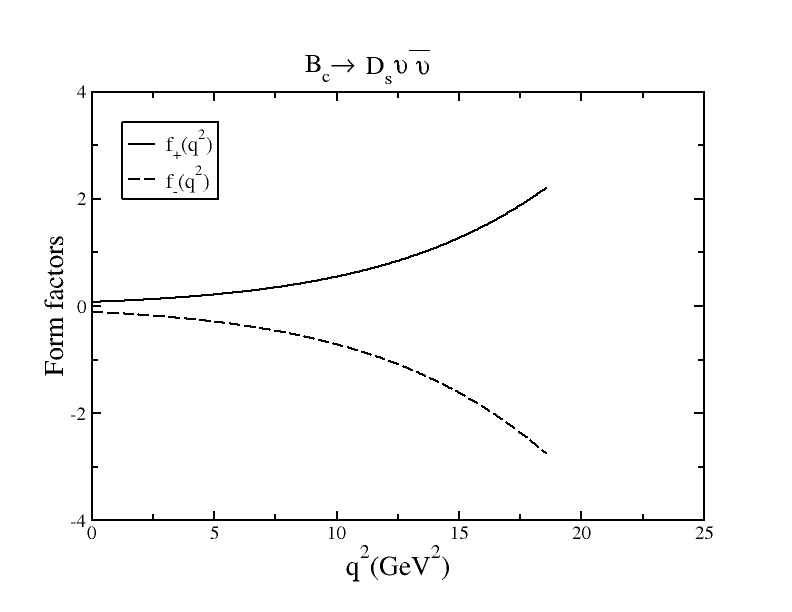}
	\includegraphics[width=0.6\textwidth]{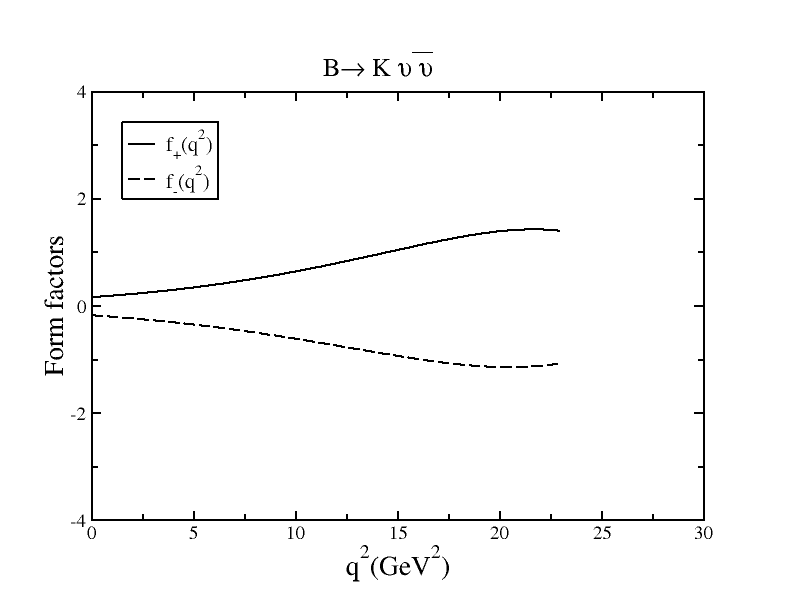}
	\caption{$q^2$-dependence of the form factors.}
\end{figure}

\begin{figure}[!hbt]
	\centering
	\includegraphics[width=0.6\textwidth]{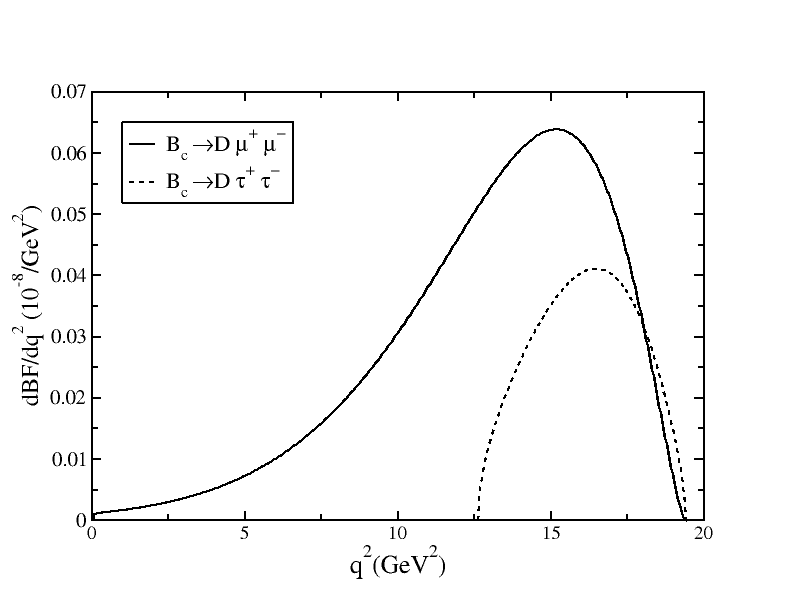}
         \includegraphics[width=0.6\textwidth]{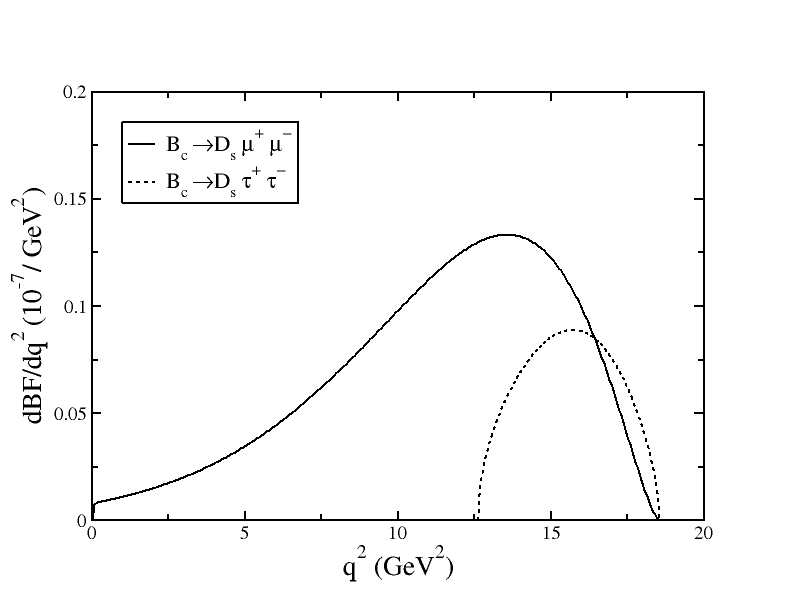}
	\includegraphics[width=0.6\textwidth]{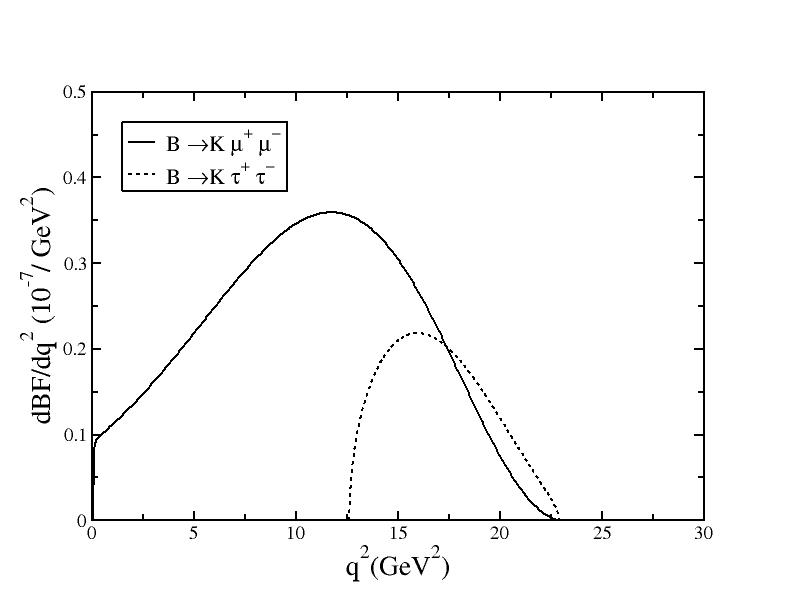}
 \caption{Predicted differential branching fractions for $B_{(c)}\to X \mu^+\mu^-$ (solid line) and $B_{(c)}\to X \tau^+\tau^-$ (dotted line), $X= D, D_s, K$.}
\end{figure}
\begin{figure}[!hbt]
	\centering
	\includegraphics[width=0.6\textwidth]{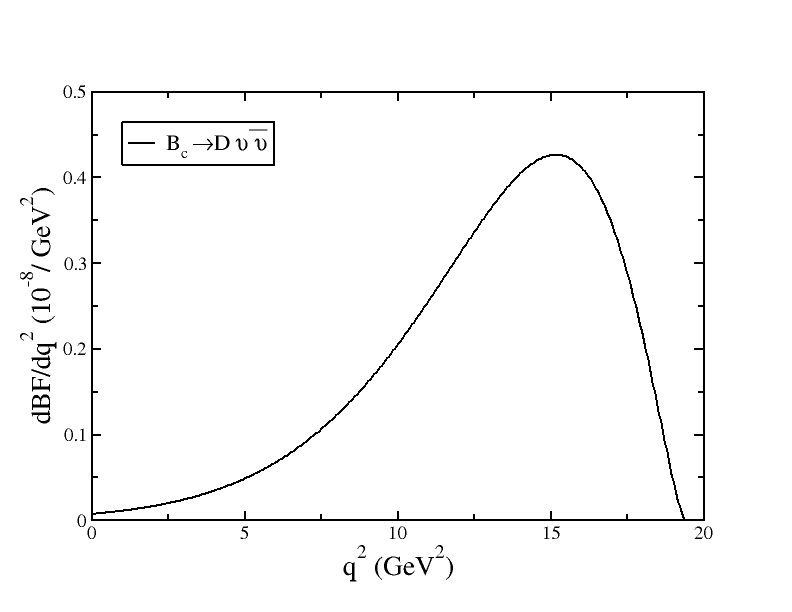}
	\includegraphics[width=0.6\textwidth]{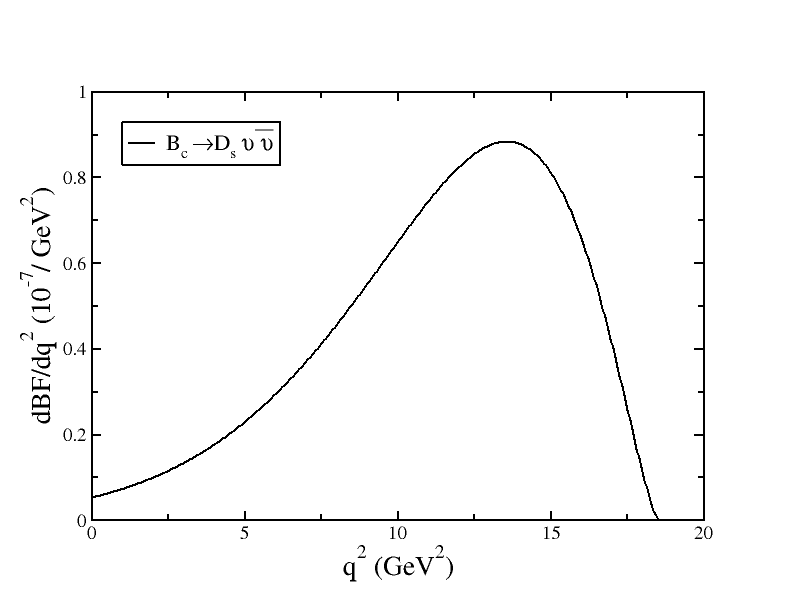}
	\includegraphics[width=0.6\textwidth]{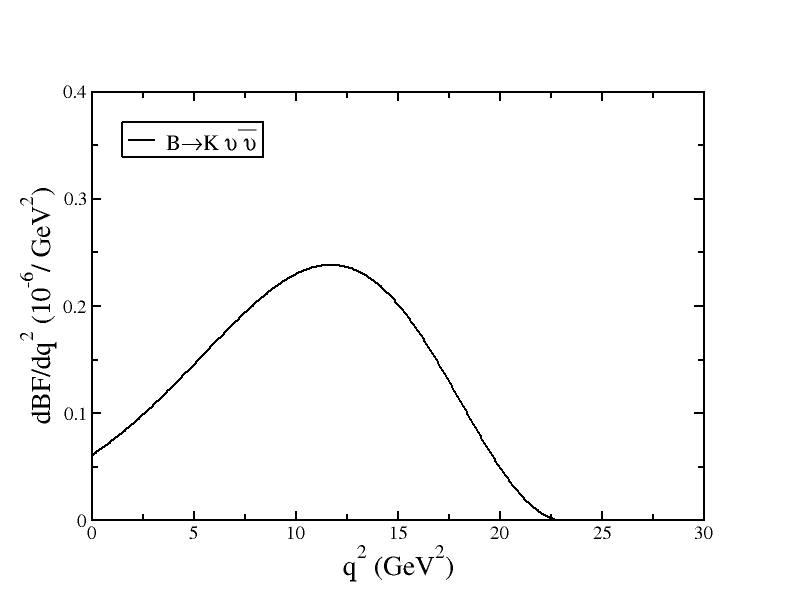}
	\caption{Predicted differential branching fractions for $B_{(c)}\to X \Sigma{\nu_l}{\bar{\nu_l}}$, $X= D, D_s, K$.}
 \end{figure}
 \begin{figure}[!hbt]
	\centering
	\includegraphics[width=0.85\textwidth]{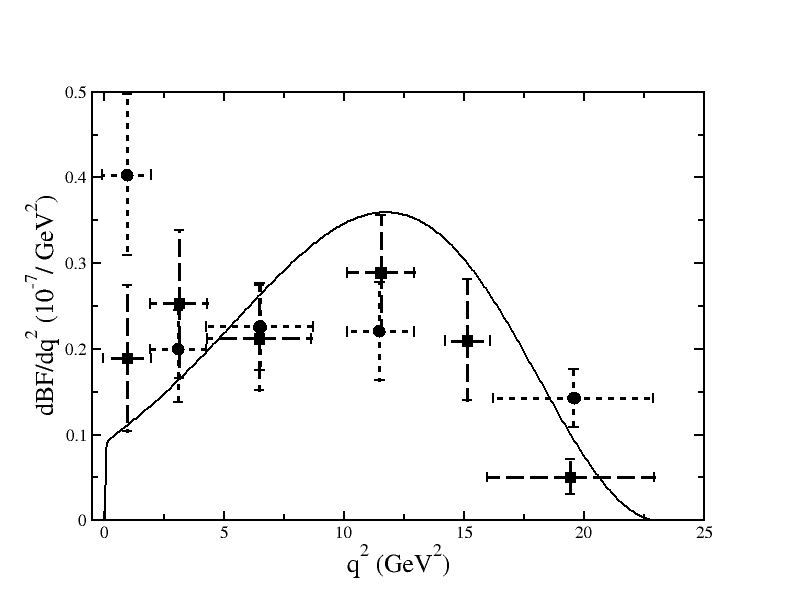}
	\caption{Comparison of our prediction for the differential branching fractions $B\to K\mu^+\mu^-$ decay with available experimental data. Belle data are given by filled circles with dotted error bars and CDF data are presented by filled squares with dashed error bars.}
\end{figure}
\begin{figure}[!hbt]
	\centering
	\includegraphics[width=0.5\textwidth]{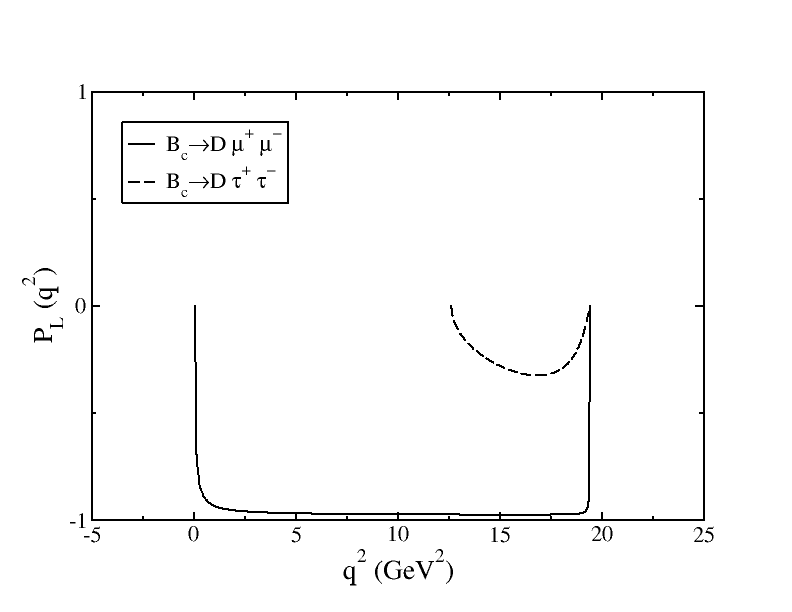}
	\includegraphics[width=0.5\textwidth]{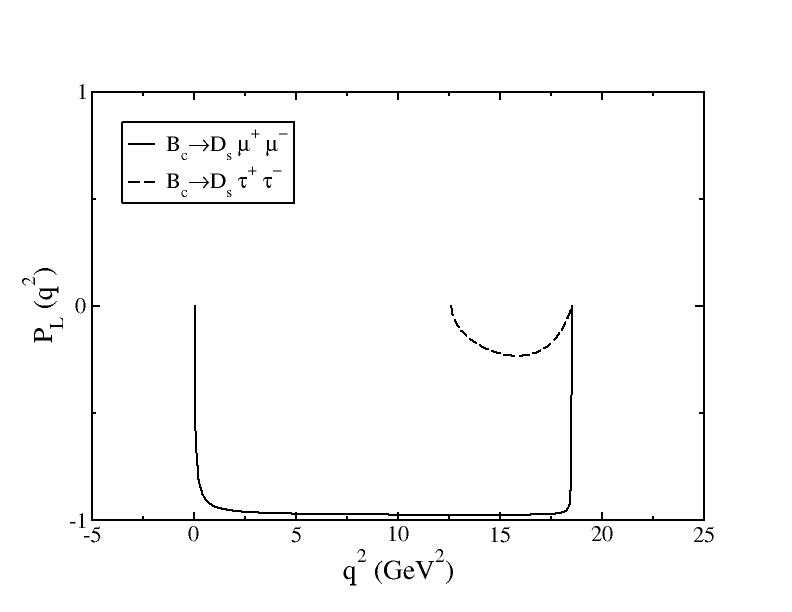}
	\includegraphics[width=0.5\textwidth]{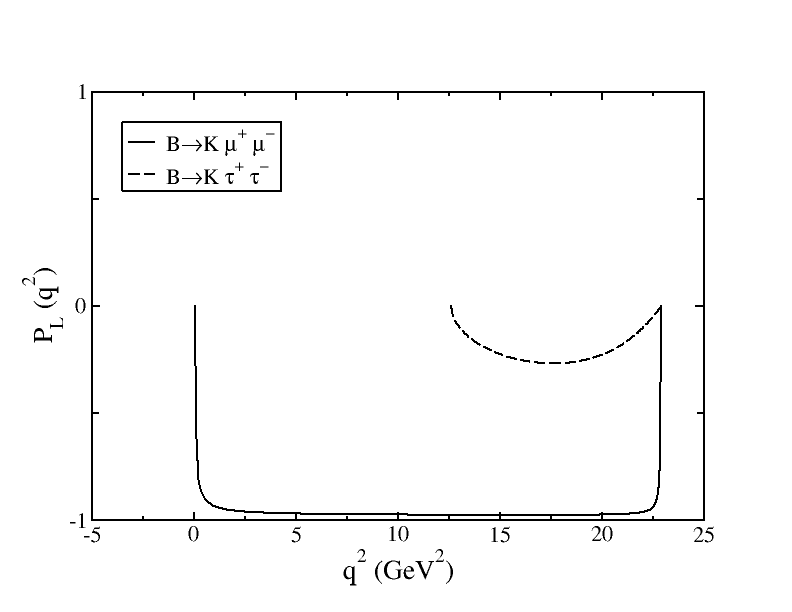}
	\caption{Predicted longitudinal lepton polarization asymmetries for $B(B_c)\to K(D_{(s)}) \mu^+\mu^-$ (solid line) and $B(B_c)\to K(D_{(s)})\tau^+\tau^-$ (dashed line).}
\end{figure}
 \par The form factors at the maximum recoil point (i.e $q^2=q^2_{min}$) correspond to the final state meson recoiling with maximum three momentum $|\vec{k}|=M^2-m^2/2M$ in the rest frame of the decaying $B_{(c)}$ meson, whereas at the zero recoil point (i.e $q^2=q^2_{max}$) they correspond to the overlap integral of the initial and final state meson wave functions. Our prediction on the values of the form factors ($f_+(q^2)$, $f_-(q^2)$ and $f_T(q^2)$) at $q^2\to q^2_{min}$ and $q^2\to q^2_{max}$ for the decay modes $B(B_c)\to K(D_{(s)})\mu^+\mu^-/ \Sigma\nu_l\bar{\nu_l}$ are shown in Table III and IV, respectively. Our results on the form factors in $B(B_c)\to K(D_{(s)})\mu^+\mu^-/ \Sigma\nu_l\bar{\nu_l}$ at $q^2\to q^2_{min}$ are comparable to those of Ref. \cite{ebert2010rare} in their calculation in the framework of the QCD-motivated relativistic quark model based on the quasi potential approach and Ref. \cite{choi2010light} in their calculation based on the light front quark model using coulomb plus harmonic oscillator (HO) and coulomb plus linear confining potential. However, our predicted form factors at $q^2=q^2_{max}$ in the above decay process along with those in the decays: $B\to K\mu^+\mu^-/\Sigma\nu_l\bar{\nu_l}$ at $q^2_{min}$ and $q^2_{max}$ are found somewhat underestimated compared to the results of Ref. \cite{ebert2010rare, choi2010light} and those of other SM predictions based on the RCQM \cite{faessler2002exclusive}, LF and CQM \cite{PhysRevD.65.094037} and QCD SR \cite{azizi2008analysis} approach. The upcoming experimental probes planned at Tevatron and LHCb may distinguish these different model predictions. Our predicted form factors of the decays $B({B_c})\to K(D_{(s)})\tau^+\tau^-$ at $q^2_{min}$ and $q^2_{max}$ are shown in Table V, which can be tested in future experiments.

\begin{table}[!hbt]
	\renewcommand{\arraystretch}{0.75}
	\centering
	\setlength\tabcolsep{2pt}
	\caption{Results for form factors at $q^2\to q^2_{min}$ and $q^2\to q^2_{max}$ of $B({B_c})\to K(D_{(s)})\tau^+\tau^-$ decays}
	\label{tab5}
\begin{tabular}{llccc}
\hline Decay mode \ \ &Valued at&\ \ $f_+(q^2)$&\ \ $f_-(q^2)$&\ \ $f_T(q^2)$ \\
		
    \hline
  	&$q^2=q^2_{min}$&0.763&-1.089&-1.202\\
  ${B_c}\to D\tau^+\tau^-$&&&&\\
		  &$q^2=q^2_{max}$&2.755&-3.807&-4.258\\
    
  \hline
		&$q^2=q^2_{min}$&0.863&-1.110&-1.296\\
  $B_c \to D_s\tau^+\tau^-$&&&&\\
		&$q^2=q^2_{max}$&3.924&-2.749&-3.250\\

  \hline
		&$q^2=q^2_{min}$&0.849&-0.779&-0.891\\
 $B\to K \tau^+\tau^-$&&&&\\
		&$q^2=q^2_{max}$&1.408&-1.074&-1.357\\
  \hline
\end{tabular}
\end{table}
 \begin{table}[!hbt]
	\renewcommand{\arraystretch}{1.5}
	\centering
	\setlength\tabcolsep{1pt}
	\caption{Comparison of our predictions of BFs of $B\to Kl\bar{l}/\Sigma\nu_l\bar{\nu_l}$ decays with experimental data (in order $10^{-7}$)}
	\label{tab6}
	\resizebox{\textwidth}{!}
	{
		\begin{tabular}{|l|c|c|c|c|c|c|}
			\hline
			Modes& This work&Belle \cite{wei2009measurement}&BABAR \cite{aubert2006measurements, lees2017search}&CDF \cite{aaltonen2009search, Pueschel:2010rm}&LHCb \cite{aaij2014test}& Belle II \cite{adachi2024evidence}\\
			\hline
			
			$B\to K \mu^+\mu^-$&$4.968^{+0.761}_{-0.723}$&$4.8^{+0.5}_{-0.4}\pm0.3$&$3.4\pm0.7\pm0.2$&$5.9\pm1.5\pm0.4$& \ $1.56^{+0.19+0.06}_{-0.15-0.04}$&\\
			$B\to K \tau^+\tau^-$&$1.494^{+0.095}_{-0.093}$&&$<2.25\times10^{4}$&&&\\
				$B\to K\Sigma\nu_l{\bar{\nu_l}}$&$33.012^{+4.6223}_{-4.4914}$&&&&&$(2.3\pm 0.5^{+0.5}_{-0.4})\times10^{2}$\\
			\hline
			
		\end{tabular}
	}
\end{table}
 \par Now using the expressions (39, 40) of the form factors $f_+(q^2)$, $f_-(q^2)$ and $f_T(q^2)$, the differential decay rates (22, 24) for $B(B_{c})\to K(D_{(s)})l^+l^-$ and $B(B_{c})\to K(D_{(s)})\Sigma\nu_l\bar{\nu_l}$ are obtained in terms of the model quantities. Before calculating numerically the physical quantities of interest: BFs, and LPA, we study the $q^2$-dependence of the differential branching fractions of $B_{(c)}\to X\mu^+\mu^-/\tau^+\tau^-$ and $B_{(c)}\to X\Sigma\nu_l\bar{\nu_l}$ over the accessible kinematical range, where $X=D, D_s, K$. This is depicted in Fig.3 and Fig.4. Then using the input parameters (Table I) and relevant phenomenological input parameters (Table II) taken from PDG \cite{navas2024review}, we perform the numerical calculation of the BFs for different rare semileptonic decays. The uncertainties in our predicted values of BFs are obtained by varying all input parameters simultaneously within $\pm10\%$ of the central values and taking the largest variation of the BFs. Our results are shown in Tables (VI and VII) in comparison with the available experimental data \cite{wei2009measurement, aubert2006measurements, lees2017search, aaltonen2009search, Pueschel:2010rm, aaij2014test, adachi2024evidence} and other SM predictions \cite{faessler2002exclusive, ebert2010rare, choi2010light, azizi2008analysis}, and LQCD results \cite{cooper2022form, parrott2023erratum}, cited in the literature.
  \begin{table}[!hbt]
	\renewcommand{\arraystretch}{1}
	\centering
	\setlength\tabcolsep{2pt}
	\caption{Branching fractions in order $10^{-7}$ in comparison with other SM predictions}
	\label{tab7}
	\resizebox{\textwidth}{!}
	{
		\begin{tabular}{|l|c|c|c|c|c|c|c|}
			\hline
			Modes& This work&\cite{faessler2002exclusive}& \cite{ebert2010rare}&HO(LO)\cite{choi2010light}& \cite{azizi2008analysis}&  \cite{cooper2022form}& \cite{parrott2023erratum}\\
			\hline
			$B_c\to D \mu^+\mu^-$&$0.053^{+0.011}_{-0.009}$&0.038&0.037&0.011(0.018)&$0.031\pm 0.006$&...&...\\
			
			$B_c\to D_s \mu^+\mu^-$&$1.259^{+0.242}_{-0.218}$&0.86&1.16&$0.51(0.54)$&$0.61\pm0.15$&$1.00\pm0.11$&...\\
			
			$B\to K \mu^+\mu^-$&$4.968^{+0.761}_{-0.723}$&5.1&4.19&...&...&...&$7.03\pm 0.55$\\
   $B_c\to D \tau^+\tau^-$&$0.019^{+0.002}_{-0.002}$&0.009&0.015&0.004(0.008)&$0.013\pm 0.003$&...&...\\
   $B_c\to D_s \tau^+\tau^-$&$0.389^{+0.033}_{-0.032}$&0.18&0.33&0.13(0.14)&$0.23\pm 0.05$&$0.245\pm0.018$&...\\
			$B\to K \tau^+\tau^-$&$1.494^{+0.095}_{-0.093}$&0.87&1.17&...&...&...&$1.83\pm 0.13$\\
   $B_c\to D \Sigma\nu_l{\bar{\nu_l}}$&$0.356^{+0.067}_{-0.061}$&0.328&0.216&0.081(0.131)&$0.384\pm0.071$&...&...\\
			$B_c\to D_s \Sigma\nu_l{\bar{\nu_l}}$&$8.363^{+1.488}_{-1.364}$&7.3&6.5&3.7(3.9)&$4.9\pm1.2$&...&...\\
				$B\to K\Sigma\nu_l{\bar{\nu_l}}$&$33.012^{+4.622}_{-4.491}$&41.9&26.1&...&...&...&$55.8\pm3.7$\\
			\hline
			
		\end{tabular}
	}
\end{table}
\par We first compare our predicted BFs of $B\to K\mu^+\mu^-$ and $B\to K\Sigma\nu_l\bar{\nu_l}$ with the available experimental data, as cited in Table VI. Our result for $B\to K\mu^+\mu^-$ compares well with the available experimental data from Belle \cite{wei2009measurement} and CDF \cite{aaltonen2009search, Pueschel:2010rm} Collaboration. A stringent test of our prediction on the differential decay distribution with the available observed data is demonstrated in Fig.5, where we confront our predictions with the experimental data. Our result for $B\to K\tau^+\tau^-$ is found to lie within the experimental upper limit of BABAR \cite{lees2017search} and that for $B\to K\Sigma\nu_l\bar{\nu_l}$ is obtained one order of magnitude lower than the recently observed data of Belle II \cite{adachi2024evidence}. However, compared with the SM predictions, our results for $B\to K\mu^+\mu^-/\tau^+\tau^-/\Sigma\nu_l\bar{\nu_l}$ are in reasonable agreement with those of Ref. \cite{faessler2002exclusive, ebert2010rare}, Ref. \cite{ebert2010rare, parrott2023erratum} and Ref. \cite{ebert2010rare, parrott2023erratum}, respectively.
\par In the rare semileptonic $B_{c}$ meson decays, where observed data are yet to be made available, we compare our results with other SM predictions and LQCD results as shown in Table VII. For decays: $B_c\to D_{s} \mu^+\mu^-$, $B_c\to D_{(s)}\tau^+\tau^-$ and $B_c\to D_{(s)}\Sigma\nu_l\bar{\nu_l}$, our predicted BFs are found in good agreement with those of Ref. \cite{faessler2002exclusive, ebert2010rare}, Ref. \cite{ebert2010rare} and Ref. \cite{ebert2010rare, azizi2008analysis}, respectively. Our results on BFs of $B_c\to D\mu^+\mu^-$, although consistent in the same order of magnitude, it is found to be somewhat over-estimated compared to those of Ref. \cite{faessler2002exclusive, ebert2010rare, azizi2008analysis} by a factor of about $\sim 1.5$. Finally, for $B\to K\Sigma\nu_l{\bar{\nu_l}}$, our result on BF is obtained comparable to those of Refs. \cite{faessler2002exclusive, ebert2010rare, parrott2023erratum}.
\begin{table}[!hbt]
	\renewcommand{\arraystretch}{1}
	\centering
	\setlength\tabcolsep{2pt}
	\caption{Longitudinal lepton polarization asymmetry}
	\label{tab8}
		\begin{tabular}{|l|c|}
			\hline
			Modes& $\langle P_L\rangle$\\
			\hline
                $B_c\to D \mu^+\mu^-$&-0.972\\
			$B_c\to D_s \mu^+\mu^-$&-0.972\\
                $B\to K \mu^+\mu^-$&-0.970\\
                $B_c\to D \tau^+\tau^-$&-0.275\\
			$B_c\to D_s \tau^+\tau^-$&-0.194\\
			$B\to K \tau^+\tau^-$&-0.224\\
			\hline
		\end{tabular}
\end{table}
\par Finally, before calculating the LPAs in different decay modes analysed in the present study, we show their $q^2$-dependence in Fig.6 for $B(B_c)\to K(D_{(s)})\mu^+\mu^-$ in solid lines and for $B(B_c)\to K(D_{(s)})\tau^+\tau^-$ in dashed lines in their allowed kinematical range. Here, the upper panels in the figure refer to the decays to $\tau^+\tau^-$ and lower panels to the decays to $\mu^+\mu^-$ mode. We note that for all the decay to $\mu^+\mu^-$ mode considered here, the LPAs ($P_L$) are zero near the endpoints ($q^2\to q^2_{min}$ and $q^2\to q^2_{max}$) and approximate to -1 away from the endpoints. In fact, the $P_L$ for the muon decay is insensitive to the form factors e.g. our $P_L$ away from the endpoint regions is well approximated by \cite{choi2010light, roberts1996hqet, burdman1995testing}
\begin{equation}
    P_L\simeq2\frac{C_{10}ReC_9^{eff}}{|C_9^{eff}|^2+|C_{10}|^2}\simeq-1
\end{equation}
in the limit of $C_7^{eff}\to 0$. It also shows that the $P_L$ for the $\mu$ dilepton channel is insensitive to the little variation of $C_7^{eff}$, as expected. On the other hand, the $P_L$ for the $\tau$ dilepton channel is somewhat sensitive to the form factors.
\par Our calculated values of the averaged LPAs are shown in Table VIII. Averaged LPA $(\langle P_L\rangle)$ for decays to $\mu^+\mu^-$ modes are obtained close to -1 i.e $\langle P_L(B_c\to D_{(s)}\mu^+\mu^-)\rangle$=-0.972 and $\langle P_L(B\to K\mu^+\mu^-)\rangle$=-0.97. However, for decays to corresponding $\tau^+\tau^-$ mode, we obtain $\langle P_L(B_c\to D\tau^+\tau^-)\rangle$=-0.275, $\langle P_L(B_c\to D_s\tau^+\tau^-)\rangle$=-0.194 and $\langle P_L(B\to K\tau^+\tau^-)\rangle$=-0.224.
\section{summary and conclusion}
In this work, we investigate the rare semileptonic  $B(B_c)\to K(D_{(s)})l\bar{l}/\Sigma\nu_l\bar{\nu_l}$ ($l=\mu, \tau$) decays in the framework of the RIQ model based on the average flavor independent confining potential in equally mixed scalar-vector harmonic form. The weak form factors ($f_{\pm}(q^2), f_T(q^2)$), expressed through the overlap integrals of the initial and final meson wave functions, are obtained in the whole accessible kinematical range: $q^2_{min}\leq0\leq q^2_{max}$, without using any {\it ad hoc} assumptions and extrapolations. The meson wave functions, obtained previously in the investigations of the meson mass spectra and several decay properties of hadrons in this model, are used for the present numerical calculations. On the basis of the form factors, the differential decay distributions and branching fractions for the above rare semileptonic $B_{(c)}$ meson decays are predicted.
\par First we tested our model by confronting its predictions for $B\to K\mu^+\mu^-$ decays with the available experimental data from Belle and CDF Collaboration; where our results are found in reasonable agreement with their data. For $B\to K\Sigma\nu_l\bar{\nu_l}$, our prediction on BF is found to be one order lower than the observed data of Belle II Collaboration. Our prediction on BFs for $B\to K\mu^+\mu^-$ and  $B\to K\Sigma\nu_l\bar{\nu_l}$ are found in agreement with other SM predictions and LQCD results.
\par Next, we provide our numerical results on the BFs for different decay modes. Our predicted BFs for $B\to K\mu^+\mu^-/\tau^+\tau^-(\Sigma\nu_l\bar{\nu_l})$, $B_c\to D(D_s)\mu^+\mu^-$, $B_c\to D(D_s)\tau^+\tau^-$ and $B_c\to D(D_s)\Sigma\nu_l\bar{\nu_l}$, obtained in order of $10^{-7}(10^{-6})$, $10^{-9}(10^{-7})$, $10^{-9}(10^{-8})$ and $10^{-8}(10^{-7})$, respectively, are in reasonable agreement with other SM predictions and LQCD results. For longitudinal lepton polarization asymmetries (LPAs), which provide valuable information on the flavor changing loop effects in the SM, our predicted averaged values are: $\langle P_L(B\to K\mu^+\mu^-)\rangle$=-0.97, $\langle P_L(B_c\to D_{(s)}\mu^+\mu^-)\rangle$=-0.972, $\langle P_L(B\to K\tau^+\tau^-)\rangle$=-0.224, $\langle P_L(B_c\to D\tau^+\tau^-)\rangle$=-0.275 and $\langle P_L(B_c\to D_s\tau^+\tau^-)\rangle$=-0.194.
\begin{acknowledgements}
The library and computational facilities provided by authorities of Siksha 'O' Anusandhan Deemed to be University, Bhubaneswar, 751030, India are duly acknowledged.
\end{acknowledgements}
\appendix
\section{Quark orbitals and momentum probability amplitude of the constituent quark and antiquark in RIQ model framework}\label{app}
In the RIQ model, a meson is picturized as a color-singlet assembly of a quark and an antiquark independently confined by an effective and average flavor-independent potential in the form:
\begin{equation}
 U(r)=\frac{1}{2}(1+\gamma^0)V(r), \nonumber   
\end{equation}
where $V(r)=(ar^2+V_0)$ with $a>0$, is believed to provide zeroth order constituent quark dynamics inside the meson-bound state. Here ($a$, $V_0$) are the potential parameters and $r$ is the relative distance between constituent quark and antiquark inside the meson core.  It is believed that the zeroth-order quark dynamics generated by the phenomenological confining potential $U(r)$, taken in equally mixed scalar-vector harmonic form, can adequately describe the decay process. With the interaction potential $U(r)$ put into the zeroth-order quark Lagrangian density, the ensuing Dirac equation admits a static solution of positive and negative energy as:
\begin{eqnarray}
	\psi^{(+)}_{\xi}(\vec r)\;&=&\;\left(
	\begin{array}{c}
		\frac{ig_{\xi}(r)}{r} \\
		\frac{{\vec \sigma}.{\hat r}f_{\xi}(r)}{r}
	\end{array}\;\right){{\chi}_{ljm_j}}(\hat r),
	\nonumber\\
	\psi^{(-)}_{\xi}(\vec r)\;&=&\;\left(
	\begin{array}{c}
		\frac{i({\vec \sigma}.{\hat r})f_{\xi}(r)}{r}\\
		\frac{g_{\xi}(r)}{r}
	\end{array}\;\right)\ {\tilde \chi}_{ljm_j}(\hat r),
\end{eqnarray}
where $\xi=(nlj)$ represents a set of Dirac quantum numbers specifying 
the eigenmodes. $\chi_{ljm_j}(\hat r)$ and ${\tilde \chi}_{ljm_j}(\hat r)$
are the spin angular parts given by,
\begin{eqnarray}
\chi_{ljm_j}(\hat r) &=&\sum_{m_l,m_s}<lm_l\;{1\over{2}}m_s|
	jm_j>Y_l^{m_l}(\hat r)\chi^{m_s}_{\frac{1}{2}},\nonumber\\
	{\tilde \chi}_{ljm_j}(\hat r)&=&(-1)^{j+m_j-l}\chi_{lj-m_j}(\hat r).
\end{eqnarray}
With the quark binding energy parameter $E_q$ and quark mass parameter $m_q$, written in the form $E_q^{\prime}=(E_q-V_0/2)$, $m_q^{\prime}=(m_q+V_0/2)$ and $\omega_q=E_q^{\prime}+m_q^{\prime}$, one can obtain solutions to the radial equation for $g_{\xi}(r)$ and $f_{\xi}(r)$ in the form:
\begin{eqnarray}
	g_{nl}&=& {\cal N}_{nl} \Big(\frac{r}{r_{nl}}\Big)^{l+1}\exp (-r^2/2r^2_{nl})
	L_{n-1}^{l+1/2}(r^2/r^2_{nl}),\nonumber\\
	f_{nl}&=&\frac{{\cal N}_{nl}}{r_{nl}\omega_q} \Big(\frac{r}{r_{nl}}\Big)^{l}\exp (-r^2/2r^2_{nl})\nonumber\\
	&\times &\left[\Big(n+l-\frac{1}{2}\Big)L_{n-1}^{l-1/2}(r^2/r^2_{nl})
	+nL_n^{l-1/2}(r^2/r^2_{nl})\right ],
\end{eqnarray}
where $r_{nl}= (a\omega_{q})^{-1/4}$ is a state independent length parameter, ${\cal N}_{nl}$
is an overall normalization constant given by
\begin{equation}
	{\cal N}^2_{nl}=\frac{4\Gamma(n)}{\Gamma(n+l+1/2)}\frac{(\omega_{q}/r_{nl})}
	{(3E_q^{\prime}+m_q^{\prime})},
\end{equation}
and $L_{n-1}^{l+1/2}(r^2/r_{nl}^2)$ are associated Laguerre polynomials. With the radial solutions (A3), an independent quark bound-state condition for constituent quark and antiquark is obtained in the RIQ model in the form of a cubic equation:
\begin{equation}
	\sqrt{(\omega_q/a)} (E_q^{\prime}-m_q^{\prime})=(4n+2l-1).
\end{equation}
From the solution of the cubic equation (A5), the zeroth-order binding energies of the confined quark and antiquark are obtained for all possible eigenmodes. 
\par In the relativistic independent particle picture of this model, the relativistic constituent quark and antiquark are thought to move independently inside the meson bound-state $|M(\vec{p}, S_{M})\rangle$ with their momentum $\vec p_{q_1}$ and ${\vec{p}}_{q_2}$, respectively. To study the decay process that takes place in the momentum eigenstates of participating mesons, we Fourier transform the quark orbitals (A1) to momentum space and obtain the momentum probability amplitude  $G_{q_1}(\vec p_{q_1})$ of the quark and ${\tilde G}_{q_2}(\vec p_{q_2})$ of the antiquark in the following forms:\\
\noindent For ground state mesons:($n=1$,$l=0$),
\begin{eqnarray}
	G_{q_1}(\vec p_{q_1})&&={{i\pi {\cal N}_{q_1}}\over {2\alpha _{q_1}\omega _{q_1}}}
	\sqrt {{(E_{p_{q_1}}+m_{q_1})}\over {E_{p_{q_1}}}}(E_{p_{q_1}}+E_{q_1})\nonumber\\
	&&\times\exp {\Big(-{
			{\vec {p_{q_1}}}^2\over {4\alpha_{q_1}}}\Big)},\nonumber\\
	{\tilde G}_{q_2}(\vec p_{q_2})&&=-{{i\pi {\cal N}_{q_2}}\over {2\alpha _{q_2}\omega _{q_2}}}
	\sqrt {{(E_{p_{q_2}}+m_{q_2})}\over {E_{p_{q_2}}}}(E_{p_{q_2}}+E_{q_2})\nonumber\\
	&&\times\exp {\Big(-{
			{\vec {p}}_{q_2}^2\over {4\alpha_{q_2}}}\Big)}.
\end{eqnarray}

\bibliography{ref}{}

\end{document}